\documentclass[10pt,twocolumn]{article}
\usepackage[top=1.8cm, bottom=1.8cm, left=1.8cm, right=1.8cm]{geometry}
\usepackage{helvet}  %
\usepackage{courier}  %
\usepackage[hyphens]{url}  %
\usepackage[keeplastbox]{flushend}  %
\usepackage{graphicx}  %
\newif\ifcomment
 \commentfalse

\usepackage{xcolor}
\usepackage{booktabs}
\usepackage{graphicx}
\usepackage{paralist}
\usepackage[small,bf]{caption}
\usepackage{subfigure}
\usepackage{times}
\usepackage[hang,flushmargin]{footmisc}
\renewcommand{\footnotesize}{\fontsize{8}{9}\selectfont}
\usepackage[compact]{titlesec}
\titlespacing*{\section}{0pt}{*4}{4pt} 
\titlespacing{\subsection}{0pt}{*3}{3pt}

\usepackage[bookmarks=true,%
            bookmarksnumbered=true,%
            colorlinks=true,%
            linkcolor=linkcol,%
            citecolor=citecol,%
            urlcolor=urlcol,%
            hypertexnames=true,%
            pdfpagelabels]%
      {hyperref}

\definecolor{linkcol}{rgb}{0,0,0.5}
\definecolor{citecol}{rgb}{0,0.5,0.3}
\definecolor{urlcol}{rgb}{0.3,0,0}

\usepackage{xspace}

\usepackage{todonotes} %

\usepackage{ifthen}
\newif\ifshort
\shortfalse %
\ifshort
  \newcommand{\isShort}{true}
\else
  \newcommand{\isShort}{false}
\fi
\newcommand{\shortVer}[1]{\ifthenelse{\equal{\isShort}{true}}{{#1}}{}}
\newcommand{\longVer}[1]{\ifthenelse{\equal{\isShort}{false}}{{#1}}{}}

\ifcomment
\newcommand{\XXX}[2]{{\bf \textcolor{blue}{#1: #2}}}
\newcommand{\XXXR}[2]{{\bf \textcolor{red}{#1: #2}}}
\newcommand{\jbnote}[1]{{\bf \textcolor{magenta}{JB: #1}}}
\newcommand{\msnote}[1]{{\bf \textcolor{magenta}{MS: #1}}}
\newcommand{\edc}[1]{{\bf \textcolor{magenta}{EDC: #1}}}
\newcommand{\gs}[1]{{\bf \textcolor{red}{gs: #1}}}

\else
\newcommand{\XXX}[2]{}
\newcommand{\XXXR}[2]{}
\newcommand{\jbnote}[1]{}
\newcommand{\msnote}[1]{}
\newcommand{\edc}[1]{}
\newcommand{\gs}[1]{}

\fi

\newcommand{\descr}[1]{\smallskip\noindent\textbf{#1}}

\renewenvironment{thebibliography}[1]{
  \begin{oldthebibliography}{#1}
    \setlength{\itemsep}{0.1em}
    \setlength{\parskip}{0.1em}
}
{
  \end{oldthebibliography}
}

\renewcommand{\footnoterule}{%
  \kern -3pt
  \hrule width 1in 
  \kern 2pt
}

\newcommand{\dspol}{{/pol/}\xspace}

\usepackage{url}
\makeatletter
\def\url@leostyle{%
  \@ifundefined{selectfont}{\def\UrlFont{}}%
  {\def\UrlFont{}}%
}
\makeatother
\definecolor{darkred}{RGB}{153,0,0}
\definecolor{darkblue}{RGB}{0,0,119}
\hypersetup{colorlinks=true, linkcolor = darkred,   citecolor = darkred, urlcolor = black}

\newif\ifwatermark
\watermarkfalse

\ifwatermark
    \usepackage{draftwatermark}
    \SetWatermarkScale{.7}
    \SetWatermarkText{\shortstack{In Submission:\\Do Not Distribute}}
    \SetWatermarkColor[rgb]{1,0.88,0.88}
\fi

\usepackage{etoolbox}
\makeatletter
\patchcmd\@combinedblfloats{\box\@outputbox}{\unvbox\@outputbox}{}{%
   \errmessage{\noexpand\@combinedblfloats could not be patched}%
}%
 \makeatother
\AtBeginShipout{%
  \ifnum\value{page}>1 %
    \typeout{* Additional boxing of page `\thepage'}%
    \setbox\AtBeginShipoutBox=\hbox{\copy\AtBeginShipoutBox}%
  \fi
}

\begin{document}
\title{\bf Understanding Web Archiving Services\\and Their (Mis)Use on Social Media\thanks{A preliminary version of this paper appears in the Proceedings of the 12th International AAAI Conference on Web and Social Media (ICWSM 2018). This is the full version.}}
\author{Savvas Zannettou$^{\star}$, Jeremy Blackburn$^\ddagger$, Emiliano De Cristofaro$^\dagger$,\\Michael Sirivianos$^{\star}$, Gianluca Stringhini$^\dagger$\\[0.5ex]
\normalsize $^{\star}$Cyprus University of Technology, $^\dagger$University College London, ${^\ddagger}$University of Alabama at Birmingham\\
\normalsize sa.zannettou@edu.cut.ac.cy, blackburn@uab.edu, \{e.decristofaro,g.stringhini\}@ucl.ac.uk, michael.sirivianos@cut.ac.cy}

\date{}

\maketitle
\begin{abstract}
Web archiving services play an increasingly important role in today's information ecosystem, by ensuring the continuing availability of information, or by deliberately caching content that might get deleted or removed.
Among these, the Wayback Machine has been \emph{proactively} archiving, since 2001, versions of a large number of Web pages, while newer services like archive.is allow users to create \emph{on-demand} snapshots of specific Web pages, which serve as time capsules that can be shared across the Web.
In this paper, we present a large-scale analysis of Web archiving services and their use on social media, shedding light on the actors involved in this ecosystem, the content that gets archived, and how it is shared. 
  We crawl and study: 1) 21M URLs from archive.is, spanning almost two years; and 2) 356K archive.is plus 391K Wayback Machine URLs that were shared on four social networks: Reddit, Twitter, Gab, and 4chan's Politically Incorrect board (\dspol) over 14 months. 
We observe that news and social media posts are the most common types of content archived, likely due to their perceived ephemeral and/or controversial nature. Moreover, URLs of archiving services are extensively shared on ``fringe'' communities within Reddit and 4chan to preserve possibly contentious content. Lastly, we find evidence of moderators nudging or even forcing users to use archives, instead of direct links, for news sources with opposing ideologies, potentially depriving them of ad revenue.
\end{abstract}

\section{Introduction}
In today's digital society, the availability and persistence of Web resources are very relevant issues.
A substantial number of URLs shared on the Web becomes unavailable after some time as websites are shutdown or redesigned in a way that does not preserve old URLs -- a phenomenon known as \emph{``link rot''}~\cite{koehler2004longitudinal}.
Moreover, content might be taken down by authorities on a legal basis, deleted by users who have shared it on social media, removed as per the ``right to be forgotten'', etc~\cite{gdpr_right_forgotten}.
Overall, the ephemerality of Web content often prompts debate with respect to its impact on the availability of information, accountability, or even censorship. 

In this context, an important role is played by services like the Wayback Machine (\url{archive.org}), which \emph{proactively} archives large portions of the Web, allowing users to search and retrieve the history of more than 300 billion pages.
At the same time, {\em on-demand} archiving services like archive.is have also become popular: users can take a snapshot of a Web page by entering its URL, which the system crawls and archives, returning a permanent short URL serving as a time capsule that can be shared across the Web.

Archiving services  serve a variety of purposes beyond addressing link rot.
Platforms like archive.is are reportedly used to preserve controversial blogs and tweets that the author may later opt to delete~\cite{mondal2016forgetting}.
Moreover, 
they also reduce Web traffic toward ``source URLs'' when the original content is still accessible, thus depriving them of potential ad revenue streams (users do not visit the original site, but just the archived copy). 
In fact, anecdotal evidence has emerged that alt-right communities target outlets they disagree with by nudging their users to share 
archive URLs instead~\cite{motherboard_archive_traffic}, or discrediting them by pointing at earlier versions of articles~\cite{vice_banned_archive}.

Given the role in helping content persist, their use on social networks, as well as anecdotal evidence of their misuse in contexts where information could be weaponized~\cite{weaponized_nytimes}, archiving services are arguably impactful actors that should be thoroughly analyzed.
To this end, this paper aims to shed light on the Web archiving ecosystem, aiming to answer the following research questions:
How are archive URLs disseminated across popular social networks?
What kind of content gets archived, by whom and why?
Are archiving services misused in any way?

To answer these questions, we perform a large-scale quantitative analysis of Web archives, based on two data sources:
1) 21M URLs collected from the archive.is live feed, and 2) 356K archive.is plus 391K Wayback Machine URLs that were shared on four social networks: Reddit, Twitter, Gab, and 4chan's Politically Incorrect board (\dspol).
Our main findings include:%
\begin{compactenum}
\item News and social media posts are the most common types of content archived, likely due to their (perceived) ephemeral and/or controversial nature. 
\item URLs of archiving services are extensively shared on ``fringe'' communities within Reddit and 4chan to preserve possibly contentious content, or to refer to it without increasing the Web traffic to the source. We also find that \dspol and Gab users favor archive.is over Wayback Machine (respectively, 15x and 16x), highlighting a particular use case in ``controversial'' online communities.
\item Web archives are exploited by users to bypass censorship policies in some communities: for instance, \dspol users post archive.is URLs to share content from 8chan and Facebook, which are banned on the platform, or to circumvent accidental censorship 
of some news sources because of substitution filters
(e.g., `smh' becomes `baka', so links to \url{smh.com.au} are unusable).
\item Reddit bots are responsible for posting a very large portion of archive URLs in the subreddits we study (respectively, 44\% and 85\% of archive.is and Wayback Machine URLs).
This is due to moderators aiming to alleviate the effects of link rot on the platform; however, this pro-active archival of content also impacts traffic to archived sites originating from Reddit.
\item The\_Donald subreddit systematically targets ad revenue of news sources with conflicting ideologies:  moderation bots block URLs from 
those sites and prompt users to post archive URLs instead (e.g., \url{nydailynews.com} have 46\% of their 
content censored). According to our conservative estimates, popular news site like the Washington Post lose yearly approx. \textdollar 70K 
from their ad revenue because of the use of archiving services on Reddit.
\end{compactenum}

\section{Related Work}
\label{sec:related_work}

\noindent\textbf{Web archives.} \cite{alnoamany2014and} analyze 6M access logs from the Wayback Machine, aiming to understand what users 
are looking for, and why they use it. 
They find that users visit the site predominantly via referrals, and that they mostly look for English pages, 
while most popular country-specific domains are from Japan, Russia, and Germany.
\cite{alonso2017s} simulate a Web archiving service, 
studying social discourse through the URLs as well as relevant entities and metadata, by analyzing millions of 
tweets as well as a case study related to fake news.
\cite{ainsworth2011how} measure how much content is available on Web archiving services:
they sample URL shorteners and search engines, query 12 public archives, and find that 35\%-90\% of URLs have at least one archived copy.
Finally,~\cite{hale2017live} assess whether the Wayback Machine archives a purely random sample of Web pages, finding 
some bias towards more visible and prominent pages.

\descr{Archived content.} \cite{nikiforakis2012you} study the evolution of JavaScript using historical data of 3M Web pages from the Wayback Machine, 
while~\cite{lerner2016internet} analyze how Web trackers have evolved in the previous decade.
\cite{soska2014automatically} predict whether an uncompromised website will 
become malicious also using the Wayback Machine. 
\cite{holzmann2016dawn} study how the German Web has evolved in the past 18 years using data from the Wayback 
Machine for 100 popular domains, highlighting the exponential growth of the number of pages. 
Then,~\cite{hackett2004accessibility} characterize the accessibility of websites using a random sample of 
sites from the Wayback Machine between 1997 and 2002.

\descr{Security.} Researchers have also focused on the security aspects of Web archiving services and link shorteners. 
\cite{lerner2017rewriting} study and address vulnerabilities on the Wayback Machine which allow attackers to 
manipulate the archived content by injecting JavaScript code.
\cite{georgiev2016gone} find that the 5- and 6-character space of link shortneners is small 
and can be easily scanned using simple algorithms, thus, an attacker could access personal/sensitive data. 
\cite{maggi2013two} perform a two-year measurement of users' interactions with 622 
URL shorteners, showing that a small subset of the users encounter malicious URLs and content. 
Finally,~\cite{nikiforakis2014stranger} show that
ad-based shorteners are more hazardous compared to traditional ones.

\descr{Remarks.}
In contrast to previous work, our analysis focuses on how archived content is shared on social networks, 
how archiving services are used, and by whom.
To the best of our knowledge, our work is the first large-scale measurement to do so, as well as to provide an in-depth analysis of on-demand Web 
archiving services (such as archive.is).

\section{Background}\label{sec:background}
We now provide an overview of the Web archiving services, as well as the social networks studied in this paper.

\subsection{Web Archives}

\descr{archive.is} offers a free, on-demand archival service of Web pages: a user 
visits the service and enters a URL to be archived. 
It also acts as a link shortener which obfuscates the source URL, by generating a 5-character URL. 
For instance, \url{http://archive.is/HVbU} shows the snapshot of Google's homepage, archived on July, 03, 2012 at 07:03:24. 

\descr{Wayback Machine.} Launched in 2001, the Wayback Machine archives a large portion of Web content, storing periodic snapshots of various pages. 
It mainly works through a proactive crawler, which visits various sites and captures a snapshot of the content.\footnote{\url{http://crawler.archive.org/index.html}}
However, users can also trigger information archival on demand.
When a page is archived, an archive URL is created in the following format: https://web.archive.org/web/[{\em time of archival}]/[{\em source URL}].  For example, the archive URL \url{https://web.archive.org/web/20100205062719/http://www.google.com/} returns the version of Google's home page on  February 5, 2010, at 06:27:19 (UTC).
In the rest of the paper, we refer to the URLs generated by archiving services 
as {\bf\em archive URLs}, and to the archived URL as {\bf\em source URLs}. 

We opt to study the Wayback Machine and archive.is for a few reasons.
First, they are popular services: as of January 2018, their Alexa Global Rank is, resp., 300 and 2,920.
We also choose these two because of some important differences between them. The Wayback Machine is run by a  501(c)(3) non-profit organization, while  archive.is is hosted by Russian provider Hostkey (only accessible via HTTP in Russia).
Moreover, the former respects robots exclusion standards 
(even retroactively)
and gives website owners the right to request removal of  pages from the archive, while the latter only complies
(albeit inconsistently) 
with DMCA take-down requests. 
Finally,  archive.is is reportedly used in ``fringe'' Web communities within 4chan and Reddit, which are known for generating~\cite{bbc_4chan_pizzagate} and incubating~\cite{guardian_reddit} fake news, and for their influence on the information ecosystem~\cite{zannettou2017web}.

\subsection{Social Networks}

\descr{Twitter} is a micro-blogging social network where users broadcast short ``tweets'' to their followers.
Search and discussion around specific topics is facilitated by hashtags, %
while the action of ``retweeting'' rebroadcasts a tweet.

\descr{Reddit} is a social news aggregator site that lets users post URLs along with a title. 
Posts get up- or down-voted and this determines the order in which they are displayed on the site.
Communities on Reddit are primarily formed via so-called ``subreddits,'' 
i.e., forums created by users and dedicated to specific topics (e.g., /r/politics 
or /r/The\_Donald).

\descr{4chan} is an imageboard discussion forum.
We focus on the Politically Incorrect board (\dspol) -- the main board for politics and world events --
because archive.is is among the most popular domains shared on the board~\cite{hine2017kek}
--we also examined other boards like International (/int/) and Sports (/sp/) but found only 
84 and 713 posts that include archive URLs for /sp/ and /int/, respectively.
Two of the key features of 4chan are anonymity and ephemerality. 
There are no accounts on 4chan and posts are displayed by default as being authored by ``Anonymous.''
The only indication of identity is the ``flag'' attribute reporting the country from the user is
posting from (based on IP geo-location) or user-chosen ones, known as ``troll'' flags (e.g., Nazi, European, Muslim, Anarchist flags, etc.).
There is only a limited number of threads that can be active on a board (on \dspol it is 200); 
when a new thread is created, an old one is purged based on the ``bump'' system~\cite{hine2017kek}.
Although several boards have a temporary archive for purged posts, all threads are permanently deleted after 7 days.

\descr{Gab} is a social network launched in 2016 to ``champion free speech, individual liberty, and the 
free flow of information online.'' It is a hybrid of Twitter and Reddit: users can broadcast 300 character messages (called ``gabs'') to their followers, 
while a voting system determines the popularity of content. Gab has been criticized for high degree of racism~\cite{gab_racism}, 
hate~\cite{gab_hate_speech,zannettou2018gab}, and for attracting alt-right users that are banned from mainstream communities like Twitter~\cite{gab_alt_right}. 
In fact, Gab's app has been deleted from Google Play for violating hate speech policy and rejected by Apple for pornographic content.

\section{Datasets}\label{sec:dataset}

We now present our datasets as well as our data collection methodology.
We perform two crawls: 1) archive.is URLs obtained from the live feed page and 
2) Wayback Machine and archive.is URLs posted on four social networks, namely, Twitter, Reddit, Gab, and 4chan's \dspol.

\descr{archive.is live feed.} To gather a large dataset of archive.is generated URLs, we use the live feed page 
(\url{http://archive.is/livefeed/}), which provides a view of the archive based on archival time (e.g., the first page lists URLs 
archived in the previous 10 minutes). In Aug 2017, we crawl the first 100K pages of the live feed,
acquiring 45.2M URLs, archived between Oct 7, 2015 and Aug 26, 2017.

Next, we visit the archive.is URLs, and scrape the content to get the archival time and the source URL.
To avoid issues for the site operators, we throttle our crawler and do not visit all 45.2M URLs. 
Instead, we randomly sample them while ensuring temporal coverage, visiting 21.5M (48\%) archive URLs, corresponding to 20.6M unique source URLs from 5.3M unique domains.
Note that given the substantial size of our sample, which guarantees temporal coverage over almost two years, the 
resulting dataset is representative of the archive. In other words, our sampling strategy does not likely introduce substantial biases affecting our results.

\descr{Archive URLs posted on social networks.} We search for archive.is and Wayback Machine URLs on Twitter, Reddit, and \dspol, 
between Jul 1, 2016 and Aug 31, 2017, and on Gab between Aug 1, 2016--Aug 31, 2017.
We obtain the 4chan dataset from~\cite{hine2017kek}, the Gab one from~\cite{zannettou2018gab}, the Reddit one from \url{pushshift.io}, while, for Twitter, we rely on the 1\% Streaming API.\footnote{\url{https://dev.twitter.com/streaming/overview}}

Overall, the resulting dataset includes 50K posts from \dspol, 528K posts from Reddit, 7K posts from Gab, and about 9K tweets.
Note that we have some gaps due to failure of our data collection infrastructure, specifically, there are 70 and 13 days missing for
Twitter and \dspol, respectively.

\descr{Basic Statistics.} In Table~\ref{tab:dataset}, we report statistics from our archive.is live feed crawl as well as the 
crawl of archive.is and Wayback Machine URLs shared on Twitter, Reddit, \dspol, and Gab. We report the number of posts with archive URLs, 
along with the percentage over the total number of posts, as well as the number of unique archive URLs, unique source URLs, unique source domains, 
and the percentage of URLs that are filtered out. 
Specifically, besides malformed URLs, we exclude, for archive.is, URLs unreachable between Aug 29 and Oct 7, 2017, while for 
Wayback Machine those pointing to types of information other than Web pages (e.g., images, videos, software, etc.).

\begin{table}[t]
\setlength{\tabcolsep}{0.18em} %
\resizebox{1\columnwidth}{!}{
\begin{tabular}{@{}llrrrrr@{}}
\toprule
\textbf{Platform}   & \textbf{Archive} & {\textbf{\#Posts with Archive}} & {\textbf{Archive}} & {\textbf{Source}} & {\textbf{Source}} & {\textbf{Filtered}} \\ 
& & {\textbf{URLs (\%all posts)}} & {\textbf{URLs}} & {\textbf{URLs}} & {\textbf{Domains}} & \\\midrule
\textbf{Live Feed} & archive.is        &                 & 21,537,554            & 20,608,834           & 5,388,112               & -                 \\ \midrule
\textbf{Reddit}     & archive.is       & 327,050 ($2.9\hspace{-0.05cm}\cdot\hspace{-0.05cm}10^{-4}\%$) & 310,392               & 291,382              & 15,994                  & 35.70\%           \\
\textbf{}           & Wayback          & 320,379 ($2.8\hspace{-0.05cm}\cdot\hspace{-0.05cm}10^{-4}\%$) & 387,081               & 343,851              & 21,124                  & 17.20\%           \\ \midrule
\textbf{/pol/}      & archive.is       & 46,912 ($1.1\hspace{-0.05cm}\cdot\hspace{-0.05cm}10^{-3}\%$)  & 36,277                & 33,824               & 3,970                   & 4.67\%            \\
\textbf{}           & Wayback          & 3,848 ($9.7\hspace{-0.05cm}\cdot\hspace{-0.05cm}10^{-5}\%$)   & 2,325                 & 2,207                & 976                     & 83.12\%           \\ \midrule
\textbf{Gab}      & archive.is       & 6,602 ($3.4\hspace{-0.05cm}\cdot\hspace{-0.05cm}10^{-4}\%$)  & 5,943              & 5,773               & 1,300                 & 5.54\%            \\
\textbf{}           & Wayback          & 478 ($5.1\hspace{-0.05cm}\cdot\hspace{-0.05cm}10^{-5}\%$)   & 361                & 349               & 240                     & 61.18\%           \\ \midrule
\textbf{Twitter}    & archive.is       & 6,750 ($3.1\hspace{-0.05cm}\cdot\hspace{-0.05cm}10^{-6}\%$)   & 3,772                 & 3,669                & 845                     & 8.23\%            \\
\textbf{}           & Wayback          & 1,905 ($9.0\hspace{-0.05cm}\cdot\hspace{-0.05cm}10^{-7}\%$)   & 1,290                 & 1,257                & 846                     & 7.49\%            \\ \bottomrule
\end{tabular}
}
\caption{Overview of our datasets: number and percentage of posts that include archive URLs, unique number of archive URLs, 
source URLs, and source domains. We also filter URLs that are malformed, unreachable, or point to resources other than Web pages.}
\label{tab:dataset}
\end{table}

Overall, \dspol and Gab users often share Wayback Machine URLs that point to non-Web pages: around 83\% and 61\% of the total, respectively,
suggesting that archive.is is used mostly for the dissemination of Web pages,
while Wayback Machine is preferred for other content.
Also, a high percentage of malformed archive.is URLs are shared on Reddit (35\%), due to bots trying to pro-actively archive resources but failing.
From the normalized percentages, we observe that Twitter users rarely share URLs from archiving services, while Reddit users do so from both archiving services.
On \dspol and Gab, we find 15 and 16 times, respectively, more archive.is URLs than Wayback Machine 
ones.

\descr{Ethical Considerations.} Although we only collect publicly available data, we have obtained clearance by our institutional ethics review board. 
Datasets and backups are stored encrypted, and we make no attempt to harm, re-identify or de-anonymize users, following 
standard ethical  principles~\cite{rivers2014ethical}.

\begin{figure*}[t]
\center
\subfigure[archive.is live feed]{\includegraphics[width=0.1955\textwidth]{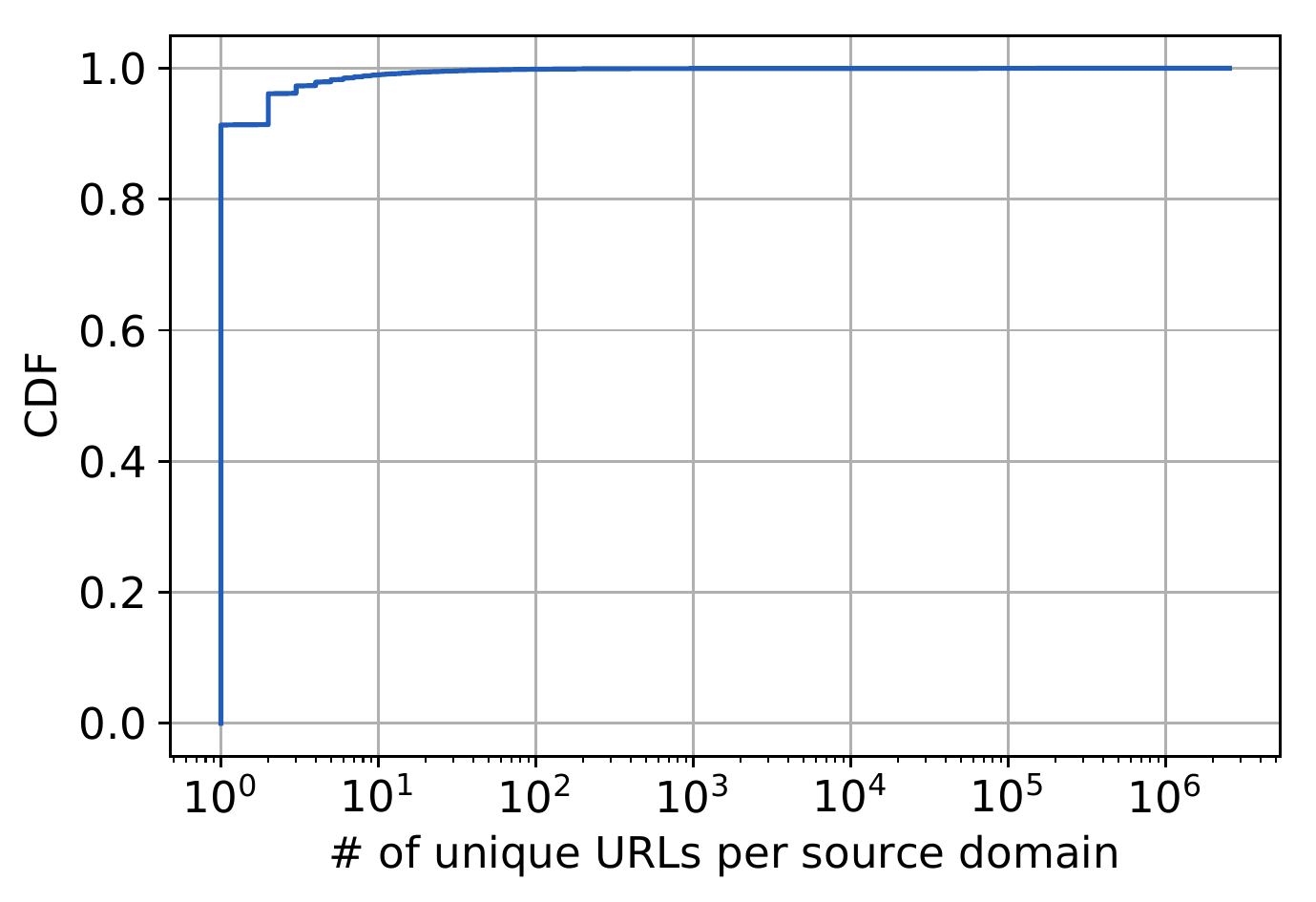}\label{fig:cdf_domain_archiveis_feed}}
\subfigure[Reddit]{\includegraphics[width=0.1955\textwidth]{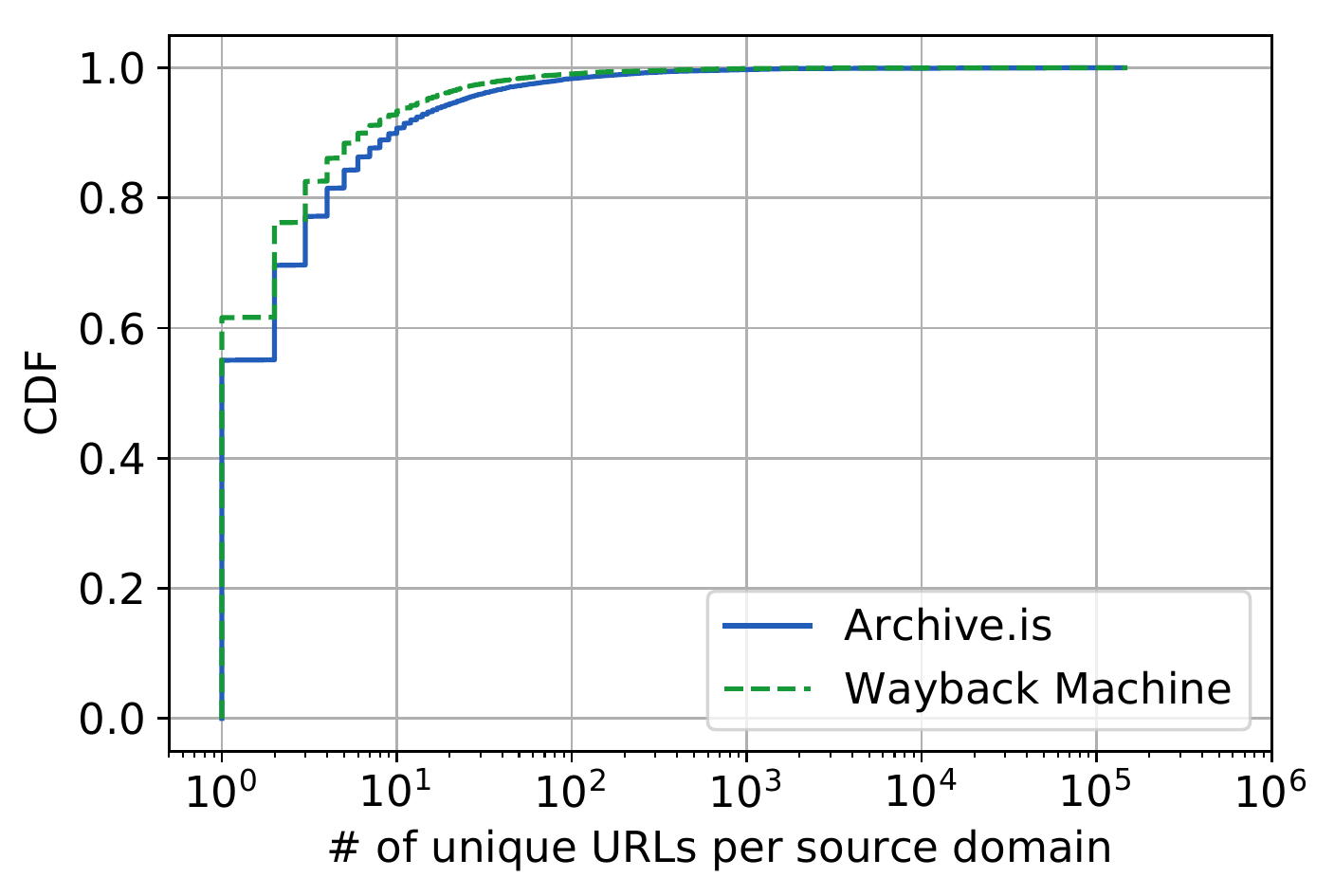}\label{subfig:cdf_domain_reddit}}
\subfigure[\dspol]{\includegraphics[width=0.1955\textwidth]{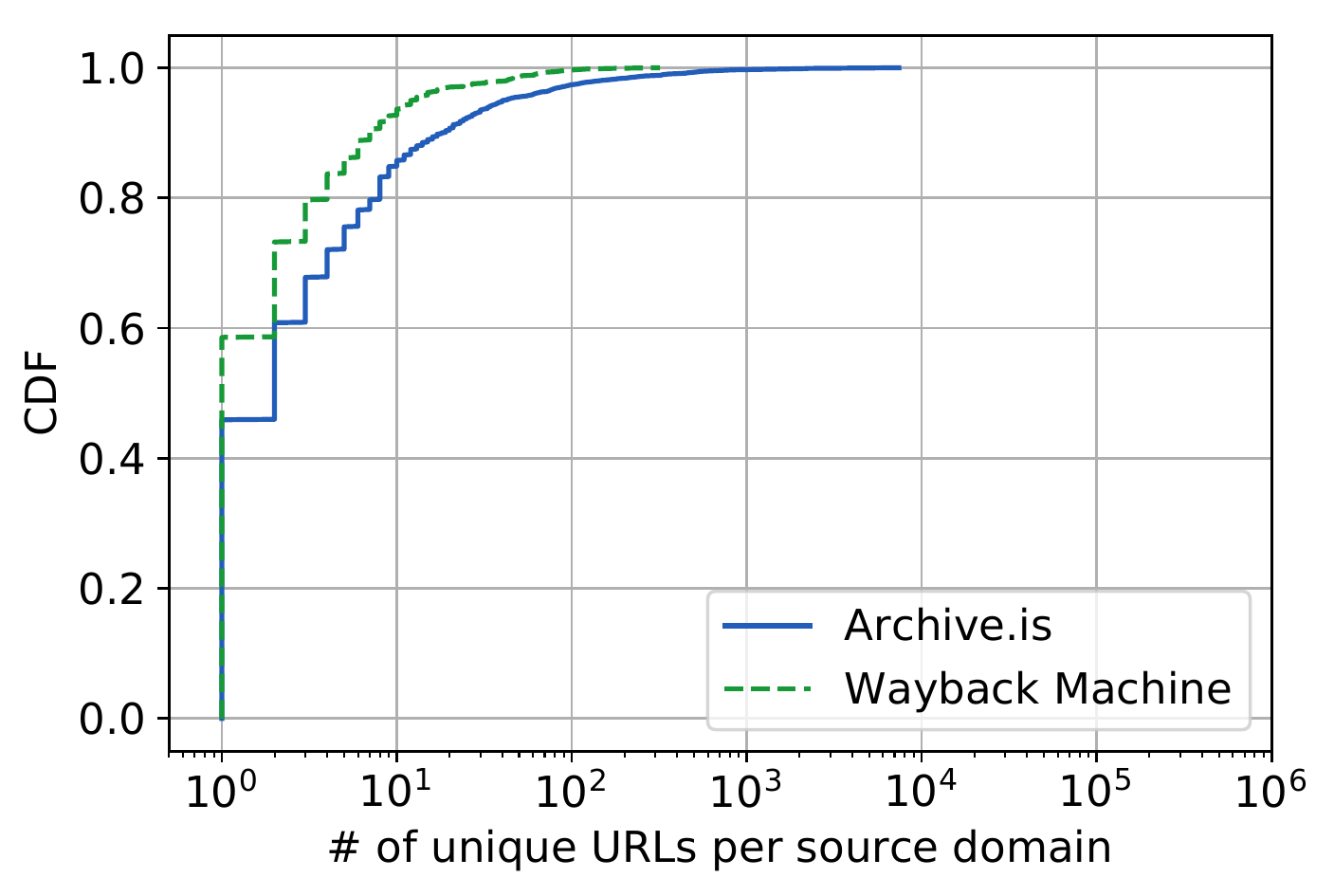}\label{subfig:cdf_domain_4chan}}
\subfigure[Twitter]{\includegraphics[width=0.1955\textwidth]{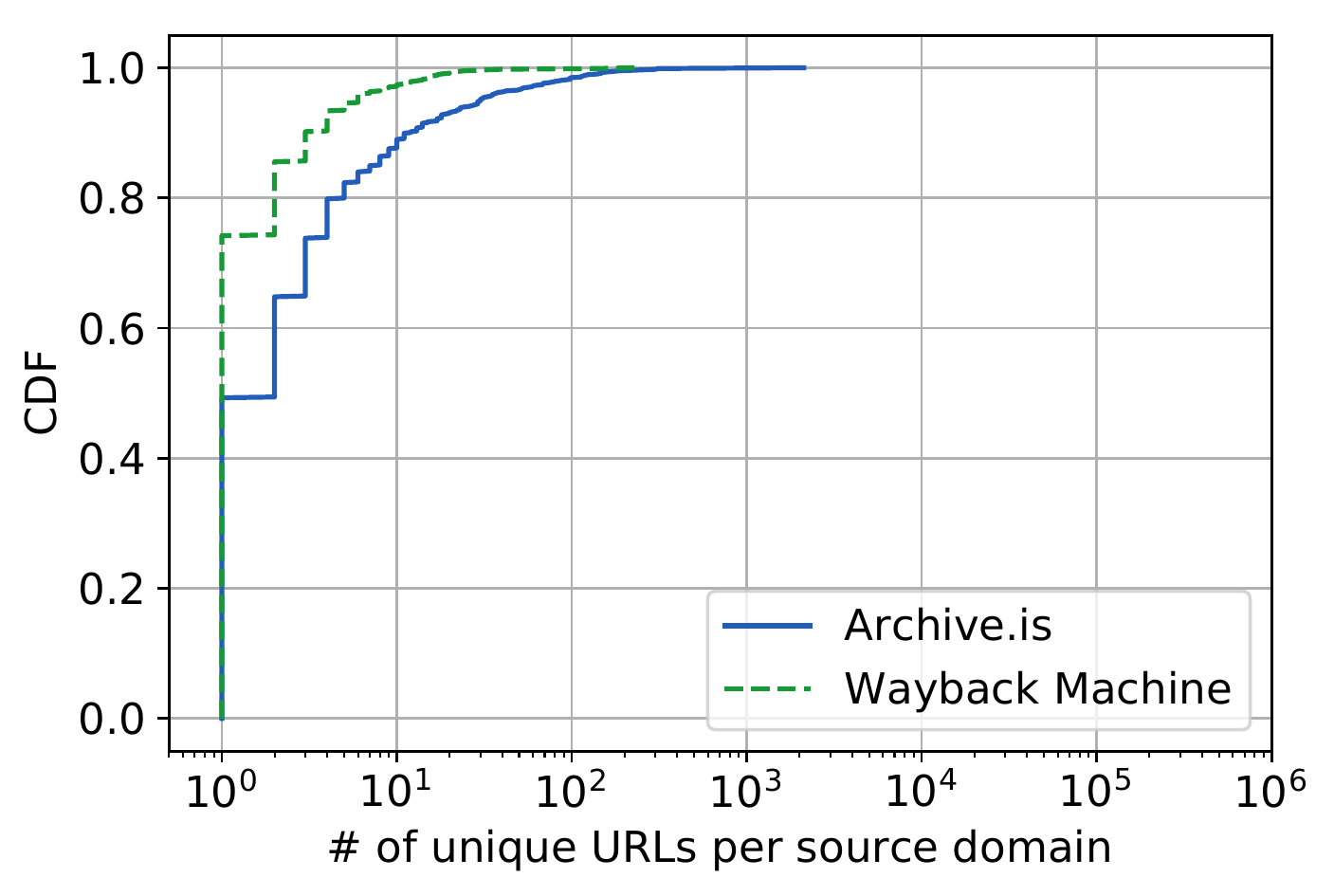}\label{subfig:cdf_domain_twitter}}
\subfigure[Gab]{\includegraphics[width=0.1955\textwidth]{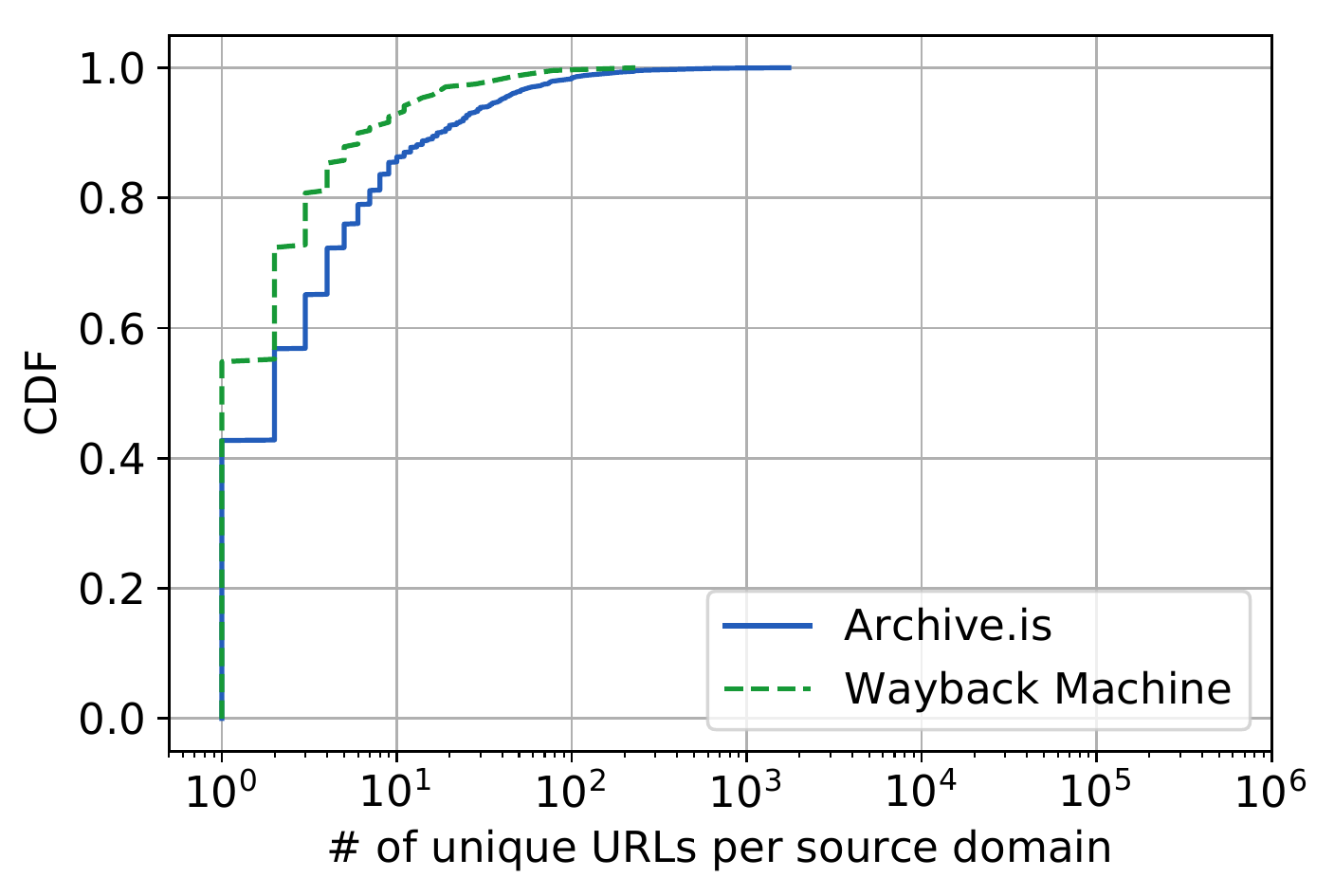}\label{subfig:cdf_domain_gab}}
\caption{CDF of the number of distinct URLs per source domain.}
\label{fig:cdf_domains}
\end{figure*}

\section{Cross-Platform Analysis}
\label{sec:analysis}

In this section, we present a cross-platform analysis of archive URLs collected from the archive.is live feed, as
well as Wayback Machine and archive.is URLs shared on the four platforms.
We focus on understanding what kind of content gets archived, the related temporal characteristics, and on 
assessing the availability of archived content from the source.

\subsection{Source Domains}

\noindent\textbf{Live Feed.} 
In Fig.~\ref{fig:cdf_domain_archiveis_feed}, we plot the CDF of the number of distinct URLs per domain in our archive.is live feed dataset.
The vast majority (90\%) of domains only appear once, while a few domains yield a large numbers of archive URLs -- e.g., there are 1.2M distinct 
archive.is URLs for which \url{twitter.com} is the source domain.
In Table~\ref{tbl:top_domains_archive.is}, we report the top 20 source domains as well as the top 20 domain suffixes (Sx). 
Surprisingly, the top domain (11.8\%) is actually the Wayback Machine's \url{archive.org}. 
Mainstream social networks like Twitter and Facebook are also included, likely due to their (perceived) ephemerality, i.e., users want to  preserve 
social network posts before they are removed or deleted.
As for the suffixes, we observe that common ones, such as .com and .org, are the majority, followed by domains from Germany 
(.de) and Japan (.jp) with 7\% and 5.6\% of the URLs, respectively. This suggests that a substantial portion of 
archive.is's user base might be in Germany and Japan.

\begin{table}[t]
\centering
\footnotesize
\setlength{\tabcolsep}{0.25em} %
\resizebox{0.99\columnwidth}{!}{%
\begin{tabular}{lrrr@{\hskip 0.12cm}|@{\hskip 0.12cm}lrrr}
\toprule
\textbf{Domain} & {\bf (\%)} & \textbf{Sx} & \multicolumn{1}{r}{\bf(\%)\hspace*{0.12cm}} & \textbf{Domain} & {\bf (\%)} & \textbf{Sx} & \textbf{(\%)} \\ \midrule
\url{archive.org}     & 11.82\% & .com & 38.29\% & \url{ru-board.com}          & 0.50\% & .pl & 1.24\% \\
\url{twitter.com}     & 5.73\%  & .org & 17.64\% & \url{asstr.org}             & 0.49\% & .ch & 1.23\% \\
\url{quora.com}       & 3.18\%  & .de  & 7.02\%  & \url{ruliweb.com}           & 0.43\% & .eu & 1.01\% \\
\url{livejournal.com} & 2.17\%  & .jp  & 5.61\%  & \url{4chan.org}             & 0.40\% & .se & 0.80\% \\
\url{reddit.com}      & 1.81\%  & .net & 3.19\%  & \url{googleusercontent.com} & 0.40\% & .cz & 0.69\% \\
\url{facebook.com}    & 1.31\%  & .ru  & 3.10\%  & \url{ameblo.jp}             & 0.39\% & .br & 0.66\% \\
\url{nhk.or.jp}       & 0.78\%  & .nl  & 2.56\%  & \url{wordpress.com}         & 0.38\% & .at & 0.63\% \\
\url{youtube.com}     & 0.65\%  & .uk  & 1.51\%  & \url{yahoo.co.jp}           & 0.38\% & .es & 0.57\% \\
\url{wikipedia.org}   & 0.52\%  & .it  & 1.39\%  & \url{aaaaarg.fail}          & 0.37\% & .be & 0.55\% \\
\url{tumblr.com}      & 0.51\%  & .fr  & 1.39\%  & \url{blogspot.nl}           & 0.36\% & .ca & 0.51\% \\ \bottomrule
\end{tabular}
}
\caption{Top 20 domains and suffixes of the source URLs in the archive.is live feed dataset.}
\label{tbl:top_domains_archive.is}
\end{table}

\descr{Social Networks. } In Figs~\ref{subfig:cdf_domain_reddit}--\ref{subfig:cdf_domain_gab}, we plot the CDF of the number of URLs for each source domain in each dataset, finding that over 40\% of the source domains only appear once. %
Wayback Machine generally archives more URLs per source domain than archive.is, although for Reddit the distributions are quite similar. %
Then, in Tables~\ref{tbl:top_domains_reddit}--\ref{tbl:top_domains_gab}, we report the top 20 source domains observed on each platform, along with their {\em archival fraction} (AF), i.e., the number of times a source domain appears in an archive over the total number of times it appears in the dataset (either archived or not).

On all platforms except for Gab, the most popular domain archived through archive.is is the platform itself; e.g., archives of tweets are the most shared ones on Twitter.
This also happens for  Wayback Machine URLs, but only on Reddit.
On Reddit, this may be due to meta-subreddits focused on the preservation and discussion of dramatic happenings, e.g., flame wars and intra-Reddit conflict, that would otherwise be lost when deleted by moderators after some time.
These meta-subreddits tend to make use of bots that automatically archive drama submitted by their members.

Overall, we notice a strong presence of both mainstream (e.g., Washington Post) and alternative (e.g., Breitbart) news sources archived and shared on Reddit, \dspol, and Gab. 
Moreover, on \dspol, archive.is is often used for links to \url{hypothes.is}, a service that lets users annotate news articles,
possibly due to the fact that \dspol users often ``unravel'' conspiracy theories by researching and commenting on news articles.
On Twitter, where the footprint of archive URLs is relatively low, we find a relatively large number of Japanese domains, which might possibly indicate a stronger presence of Japanese Twitter users relying on archives.

The AFs are quite low overall, implying that archiving services disseminate a small fraction of most domains. However, on \dspol, specific domains have extremely high AFs. For instance, we find that \url{facebook.com} (AF = 0.96) and \url{8ch.net} (AF = 1.0)
are marked as spam from \dspol, and posts including links to them are rejected, a phenomenon we refer to as \textit{platform-specific censorship}.
We manually analyze other domains with high AF values like \url{hypothes.is}, \url{chetlyzarko.com}, \url{tdbming.com}, etc., without finding evidence of censorship on \dspol.
There is also ``accidental'' censorship on \dspol: for instance, the Australian newspaper \url{smh.com.au}, is affected because of a substitution filter (used for fun), which replace one word with another, as the word ``smh'' is automatically replaced on \dspol\ with ``baka.''

\begin{table}[t]
\centering
\footnotesize
\setlength{\tabcolsep}{0.25em} %
\resizebox{0.9\columnwidth}{!}{%
\begin{tabular}{lrr|lrr}
\toprule
\textbf{Domain (archive.is)} & \textbf{(\%)} & \multicolumn{1}{r}{\bf AF} &\textbf{Domain (Wayback)} & \textbf{(\%)} & {\bf AF} \\ \midrule
reddit.com          &      31.21\%  & {$<\hspace{-0.05cm}0.01$} & reddit.com  & 36.88\%    &  $<\hspace{-0.05cm}0.01$\\
pastebin.com       &         6.80\%  & {0.08}  & imgur.com  & 7.05\%      & $<\hspace{-0.05cm}0.01$\\
twitter.com           &     5.89\%  & {$<\hspace{-0.05cm}0.01$}  & twitter.com   &  5.19\%      & $<\hspace{-0.05cm}0.01$\\
imgur.com            &  3.02\% & {$<\hspace{-0.05cm}0.01$}  & redd.it   & 4.79\%        & $<\hspace{-0.05cm}0.01$\\
washingtonpost.com &            2.46\%    & {0.02}  & youtube.com  & 3.90\%   & $<\hspace{-0.05cm}0.01$      \\
youtube.com               & 2.33\%& {$<\hspace{-0.05cm}0.01$}  & washingtonpost.com   & 1.54\%     &   0.01  \\
redd.it                   & 2.14\%& {$<\hspace{-0.05cm}0.01$}  & youtu.be   & 1.19\%          & $<\hspace{-0.05cm}0.01$\\
nytimes.com            &     1.76\% & {0.01}  & nytimes.com   & 0.98\%   &    $<\hspace{-0.05cm}0.01$  \\
cnn.com                  & 1.64\%& {0.02}  & cnn.com   & 0.90\%         &  $<\hspace{-0.05cm}0.01$\\
wikipedia.org               & 1.37\%  & {$<\hspace{-0.05cm}0.01$}  & reddituploads.com   & 0.89\%   & 0.06       \\
huffingtonpost.com        &      0.93\%  & {0.02}  & archive.is   & 0.61\%          & $<\hspace{-0.05cm}0.01$\\
theguardian.com              &    0.78\% & {$<\hspace{-0.05cm}0.01$}  & streamable.com   & 0.61\%    &  $<\hspace{-0.05cm}0.01$      \\
googleusercontent.com      &      0.65\%      & {0.08}  & thehill.com   & 0.54\%    &   0.01     \\
politico.com                   & 0.64\%& {0.02}  & wikipedia.org   & 0.52\%           &  $<\hspace{-0.05cm}0.01$\\
wsj.com                  & 0.61\%& {0.03}  & politico.com   & 0.49\%           &  0.02\\
dailymail.co.uk       & 0.54\%& {0.01}  & theguardian.com   & 0.46\%       &    $<\hspace{-0.05cm}0.01$ \\
4chan.org                 &0.53\% & {0.16}  & rawstory.com   & 0.45\%          &  0.06\\
facebook.com            &      0.52\%  & {$<\hspace{-0.05cm}0.01$}  & huffingtonpost.com   & 0.44\%  &  $<\hspace{-0.05cm}0.01$        \\
thehill.com                &  0.43\%& {0.01}  & bbc.com   & 0.44\%       &  0.01\\
breitbart.com             & 0.40\%  & {0.01}  & kickstarter.com   & 0.37\%   &     0.02           \\ \bottomrule
\end{tabular}
}
\caption{Top 20 source domains of archive.is and Wayback Machine URLs, and archival fraction (AF), in the Reddit dataset.}
\label{tbl:top_domains_reddit}
\bigskip
\centering
\footnotesize
\setlength{\tabcolsep}{0.25em} %
\resizebox{0.9\columnwidth}{!}{%
\begin{tabular}{lrr|lrr}
\toprule
\textbf{Domain (archive.is)} & \textbf{(\%)} & \multicolumn{1}{r}{\bf AF} & \textbf{Domain (Wayback)} & \textbf{(\%)} & \textbf{AF} \\ \midrule
4chan.org                & 9.35\% & \multicolumn{1}{r|}{0.54} & justice4germans.com  & 7.50\%  & \textbf{0.94}    \\
theguardian.com          & 3.78\%       & \multicolumn{1}{r|}{0.13}  & chetlyzarko.com  & 3.90\% & \textbf{1.00}    \\
washingtonpost.com       & 3.70\%         & \multicolumn{1}{r|}{0.20}  & twitter.com   &  2.82\% & $<\hspace{-0.05cm}0.01$     \\
nytimes.com             & 3.46\% & \multicolumn{1}{r|}{0.16}  & dailymail.co.uk   & 2.47\%      & $<\hspace{-0.05cm}0.01$ \\
cnn.com                 & 2.78\% & \multicolumn{1}{r|}{0.14}  & revcom.us  & 2.16\%          & 0.66 \\
twitter.com             &  2.75\%   & \multicolumn{1}{r|}{0.01}  & reddit.com   & 1.98\%       & $<\hspace{-0.05cm}0.01$   \\
independent.co.uk       &  2.37\%           & \multicolumn{1}{r|}{0.13}  & tumblr.com   & 1.85\%  & 0.02        \\
breitbart.com            & 1.96\%     & \multicolumn{1}{r|}{0.08}  & thebilzerianreport.com   & 1.57\% & 0.72         \\
reddit.com               & 1.85\%  & \multicolumn{1}{r|}{0.09}  & jeffreyepsteinscience.com   & 1.55\%  & \textbf{1.00}        \\
dailymail.co.uk          & 1.72\%       & \multicolumn{1}{r|}{0.05}  & cnn.com   & 1.51\%        & $<\hspace{-0.05cm}0.01$   \\
facebook.com             & 1.69\%  & \multicolumn{1}{r|}{\textbf{0.96}}  & tdbimg.com   & 1.43\%          & \textbf{1.00} \\
huffingtonpost.com       & 1.37\%           & \multicolumn{1}{r|}{0.20}  & huffingtonpost.com   & 1.43\%  & 0.01          \\
thehill.com              & 1.21\%   & \multicolumn{1}{r|}{0.16}  & metapedia.org   & 1.22\%          & 0.04  \\
politico.com             & 1.04\%      & \multicolumn{1}{r|}{0.13}  & nytimes.com   & 1.15\%         & $<\hspace{-0.05cm}0.01$   \\
bbc.com                  & 1.01\% & \multicolumn{1}{r|}{0.08}  & washingtonpost.com   & 1.11\%        & $<\hspace{-0.05cm}0.01$    \\
8ch.net       & 0.98\% & \multicolumn{1}{r|}{\textbf{1.00}}  & theguardian.com   & 1.08\%           &  $<\hspace{-0.05cm}0.01$\\
googleusercontent.com           &  0.91\%    & \multicolumn{1}{r|}{0.59}  & independent.co.uk   & 1.08\%  & $<\hspace{-0.05cm}0.01$          \\
hypothes.is                   & 0.87\% & \multicolumn{1}{r|}{\textbf{0.98}}  & wordpress.com   & 1.06\%           & $<\hspace{-0.05cm}0.01$ \\
telegraph.co.uk               &  0.85\% & \multicolumn{1}{r|}{0.03}  & idrsolutions.com   & 1.01\%      & 0.86   \\
theatlantic.com              &  0.81\% & \multicolumn{1}{r|}{0.24}  & wikileaks.com   & 1.01\%            & $<\hspace{-0.05cm}0.01$       \\ \bottomrule
\end{tabular}
}
\caption{Top 20 source domains of archive.is and Wayback Machine URLs, and archival fraction (AF), in the /pol/ dataset.}
\label{tbl:top_domains_4chan}
\end{table}

\begin{table}[t!]
\centering
\footnotesize
\setlength{\tabcolsep}{0.25em} %
\resizebox{0.9\columnwidth}{!}{%
\begin{tabular}{lrr|lrr}
\toprule
\textbf{Domain (archive.is)} & \textbf{(\%)} & \multicolumn{1}{r}{\bf AF} & \textbf{Domain (\url{Wayback})} & \textbf{(\%)} &\textbf{\bf AF} \\ \midrule
twitter.com                 & 25.02 \% & \multicolumn{1}{r|}{$<\hspace{-0.05cm}0.01$} & justpaste.it  & 11.90 \%    & 0.02\\
facebook.com              & 3.65 \%   & \multicolumn{1}{r|}{$<\hspace{-0.05cm}0.01$}  & twitter.com  & 6.90 \%    &  0.01\\
togetter.com                & 3.58 \% & \multicolumn{1}{r|}{$<\hspace{-0.05cm}0.01$}  & dailymail.co.uk   &  1.95 \%   &  0.13  \\
seesaa.net             & 2.97 \%& \multicolumn{1}{r|}{\textbf{0.91}}  & nikkansports.com   & 1.50 \%       &  0.18\\
justpaste.it             & 2.19 \%    & \multicolumn{1}{r|}{0.01}  & mikelofgren.net  & 1.20\%       &   \textbf{1.00}\\
yahoo.co.jp              & 2.03 \%  & \multicolumn{1}{r|}{0.21}  & blogspot.com   & 1.10\%         &  0.09\\
googleusercontent.com\hspace*{-0.2cm} & 1.77 \%                 & \multicolumn{1}{r|}{\textbf{0.98}}  & whitehouse.gov   & 1.05\%     & 0.02     \\
time.com                 & 1.75 \%& \multicolumn{1}{r|}{0.01}  & journalists-in-russia.org\hspace*{-0.2cm}   & 1.00\%         &  \textbf{1.00}\\
monjiro.net               & 1.66 \%  & \multicolumn{1}{r|}{0.51}  & pcdepot.co.jp   & 0.90\%         &  \textbf{0.90}\\
pastebin.com             &   1.45 \%  & \multicolumn{1}{r|}{0.04}  & rydon.co.uk   & 0.85\%         & \textbf{1.00}\\
google.com              & 1.39 \% & \multicolumn{1}{r|}{0.01}  & yeniakit.com.tr   & 0.85\%          &  0.16\\
jimin.jp                  & 1.35 \%& \multicolumn{1}{r|}{\textbf{0.95}}  & cdse.edu   & 0.75\%            & \textbf{0.93}\\
notepad.cc              & 1.33 \%   & \multicolumn{1}{r|}{0.47}  & tetsureki.com   & 0.75\%       &    \textbf{1.00} \\
ameblo.jp                 &  1.16 \%  & \multicolumn{1}{r|}{$<\hspace{-0.05cm}0.01$}  & donaldjtrump.com   & 0.75\%      & 0.04     \\
nhk.or.jp                  & 1.16 \% & \multicolumn{1}{r|}{0.33}  & reidreport.com   & 0.75\%            & \textbf{1.00}\\
magi.md         & 1.16 \% & \multicolumn{1}{r|}{0.49}  & ameblo.cjp   & 0.70\%           &  $<\hspace{-0.05cm}0.01$\\
opensecrets.org      &  1.05 \%         & \multicolumn{1}{r|}{0.67}  & jreast.co.jp   & 0.70\%   &    \textbf{0.93}     \\
fc2.com                   & 0.99 \% & \multicolumn{1}{r|}{0.27}  & eastandard.net   & 0.65\%       &    \textbf{1.00} \\
dailyshincho.jp          &    0.93 \%    & \multicolumn{1}{r|}{\textbf{0.94}}  & yahoo.co.jp   & 0.60\%      &   0.01\\
reddit.com              &  0.89 \% & \multicolumn{1}{r|}{0.03}  & livedoor.jp   & 0.60\%                &   0.07\\ \bottomrule
\end{tabular}
}
\caption{Top 20 source domains of archive.is and Wayback Machine URLs, and archival fraction (AF), in the Twitter dataset.}
\label{tbl:top_domains_twitter}
\bigskip
\centering
\footnotesize
\setlength{\tabcolsep}{0.25em} %
\resizebox{0.99\columnwidth}{!}{%
\begin{tabular}{lrr|lrr}
\toprule
\textbf{Domain (archive.is)} & \textbf{(\%)} & \multicolumn{1}{r}{\bf AF} & \textbf{Domain (Wayback)} & \textbf{(\%)} & \textbf{AF} \\ \midrule
twitter.com                & 12.28\% & \multicolumn{1}{r|}{$<\hspace{-0.05cm}0.01$} & dailymail.co.uk  & 20.98\%  & $ < 0.01 $    \\
nytimes.com          & 4.71\%       & \multicolumn{1}{r|}{0.03}  & washingtonpost.com  & 7.08\% & $ 0.01 $    \\
washingtonpost.com       & 4.17\%         & \multicolumn{1}{r|}{0.03}  & infowars.com   &  5.54\% & $<\hspace{-0.05cm}0.01$     \\
reddit.com            & 3.10\% & \multicolumn{1}{r|}{0.03}  & brandenburg.de   & 4.35\%      & 0.10 \\
googleusercontent.com                 & 2.43\% & \multicolumn{1}{r|}{0.18}  & twitter.com  & 3.63\%          & $ < 0.01$ \\
breitbart.com             &  1.82\%   & \multicolumn{1}{r|}{$ < 0.01$}  & huffingtonpost.com   & 3.08\%       & $<\hspace{-0.05cm}0.01$   \\
cnn.com       &  1.63\%           & \multicolumn{1}{r|}{0.01}  & abcnews.go.com   & 2.54\%  & $ < 0.01 $        \\
4chan.org            & 1.59\%     & \multicolumn{1}{r|}{0.07}  & salon.com   & 1.72\% & 0.01         \\
dailymail.co.uk               & 1.44\%  & \multicolumn{1}{r|}{$<\hspace{-0.05cm}0.01$}  & alexa.com   & 1.63\%  & 0.03        \\
theguardian.com          & 1.29\%       & \multicolumn{1}{r|}{$ < 0.01$}  & news.com.au   & 1.54\%        & $<\hspace{-0.05cm}0.01$   \\
wsj.com             & 1.22\%  & \multicolumn{1}{r|}{0.01}  & tu-dortmunt.de   & 1.45\%          & 0.80 \\
bbc.com       & 1.15\%           & \multicolumn{1}{r|}{0.01}  & causes.com   & 1.27\%  & 0.50          \\
huffingtonpost.com              & 1.14\%   & \multicolumn{1}{r|}{0.03}  & vigilantcitizen.com   & 1.18\%          & 0.02  \\
google.com             & 1.01\%      & \multicolumn{1}{r|}{$ < 0.01$}  & reddit.com   & 1.08\%         & $<\hspace{-0.05cm}0.01$   \\
facebook.com                  & 0.92\% & \multicolumn{1}{r|}{$ < 0.01 $}  & sahra-wagenknecht.de   & 0.99\%        & 0.78   \\
latimes.com       & 0.85\% & \multicolumn{1}{r|}{0.01}  & quillette.com   & 0.99\%           &  0.02\\
yahoo.com           &  0.81\%    & \multicolumn{1}{r|}{$ < 0.01$}  & derwesten.de   & 0.99\%  & $<\hspace{-0.05cm}0.01$          \\
dailycaller.com                   & 0.77\% & \multicolumn{1}{r|}{$ < 0.01$}  & politico.com   & 0.91\%           & $<\hspace{-0.05cm}0.01$ \\
thehill.com               &  0.74\% & \multicolumn{1}{r|}{$ < 0.01$}  & mikelofgren.net   & 0.81\%      & \textbf{0.90}   \\
wikileaks.org              &  0.73\% & \multicolumn{1}{r|}{0.01}  & alexanderhiggins.com   & 0.81\%            & 0.02       \\ \bottomrule
\end{tabular}
}
\caption{Top 20 source domains of archive.is and Wayback Machine URLs, and archival fraction (AF), in the Gab dataset.}
\label{tbl:top_domains_gab}
\end{table}

\subsection{URL Characterization}
We now proceed to characterize the type of archived content.
To this end, we extract the domain categories of source URLs using the free Virus Total API (\url{virustotal.com}), 
which we choose since it consolidates categories from multiple services (e.g., Bit Defender and Alexa).
Although categorization is done at domain-level, results are presented at a per-URL level (a URL is assigned the same category as its domain) to capture the popularity of each domain. %

\descr{Live Feed.} Due to throttling enforced by the API, we are not able to categorize all the 20.6M source URLs in our archive.is live feed dataset. %
Therefore, we first aggregate URLs into their domain, then, we follow a sampling approach using: 1) the top 100K most popular domains in our dataset, which correspond to 15M source URLs, and 2) a sample of 121K domains drawn according to their empirical distribution in our archive datasets, resulting in 1.4M (7\%) archive URLs. 

\begin{figure}[t]
\center
\subfigure[Top 100K Domains]{\includegraphics[width=0.495\columnwidth]{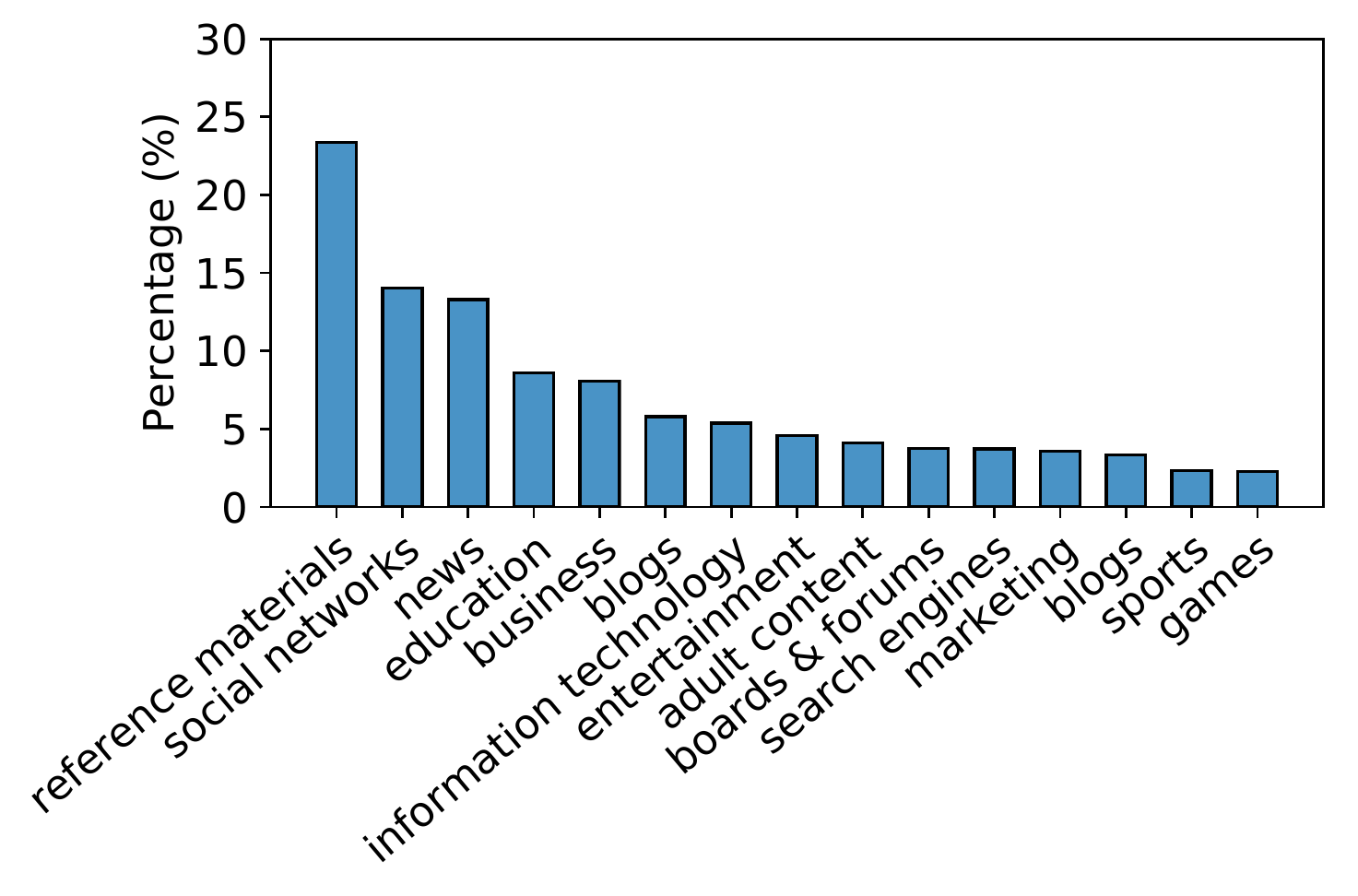}\label{subfig:bc_top_100}}
\subfigure[Sample of 121K Domains]{\includegraphics[width=0.495\columnwidth]{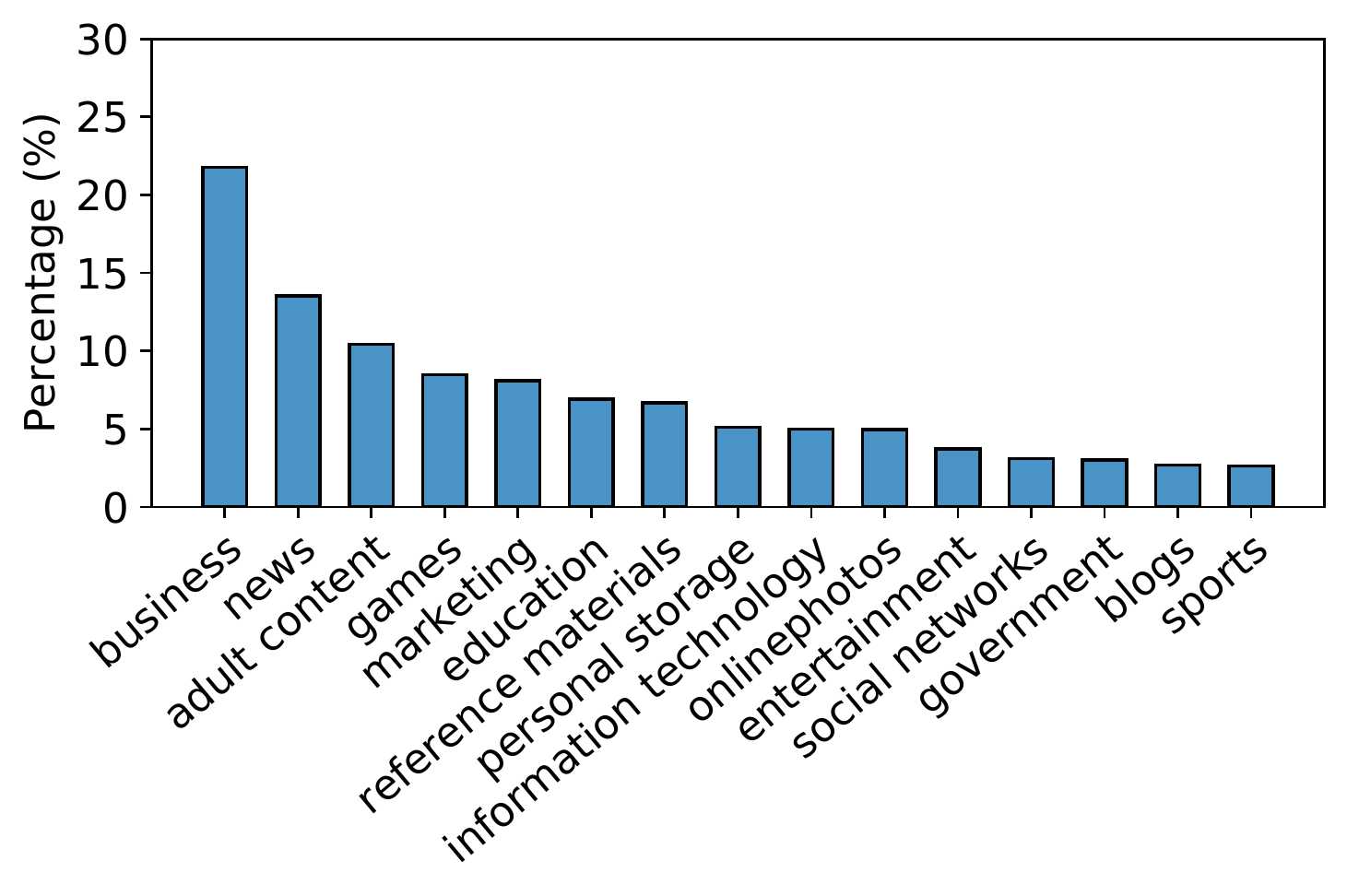}\label{subfig:bc_empirical}}
\caption{Top 15 domain categories for the archive.is live feed.}
\label{fig:bc_categorization_wild}
\end{figure}

In Fig.~\ref{fig:bc_categorization_wild}, we report the top 15 categories obtained from Virus Total for both samples.
Note that Virus Total is unable to provide a category for 1\% and 7\% of the URLs for the two sets of domains that we checked, respectively. 
From Fig.~\ref{subfig:bc_top_100}, we observe that the most popular category is Reference Materials (23\%), which is due to the fact that, as discussed earlier, many archive.is URLs archive Wayback Machine URLs.
Other popular categories include Social Networks (15\%), News Sources (14\%), Education (13\%), and Business (12\%).
Adult Content accounts for 4\% of source URLs.
Fig.~\ref{subfig:bc_empirical} shows that, for the empirically distributed sample, the top 15 categories are slightly different,
including Business (21\%), News (13\%), and Adult Content (12\%).

\begin{figure}[t]
\center
\subfigure[Reddit]{\includegraphics[width=0.495\columnwidth]{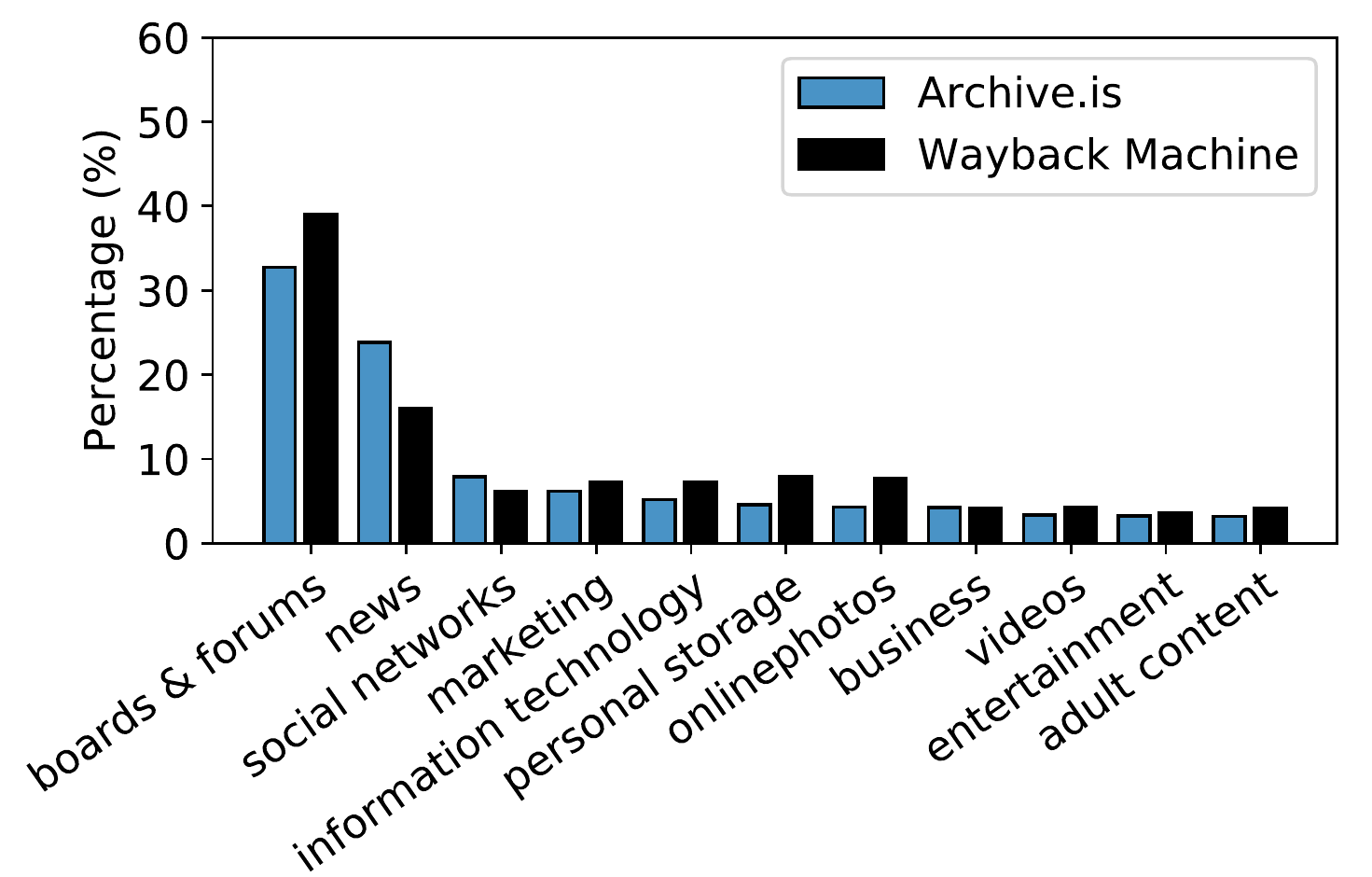}\label{subfig:bc_combined_reddit_virus}}
\subfigure[Twitter]{\includegraphics[width=0.495\columnwidth]{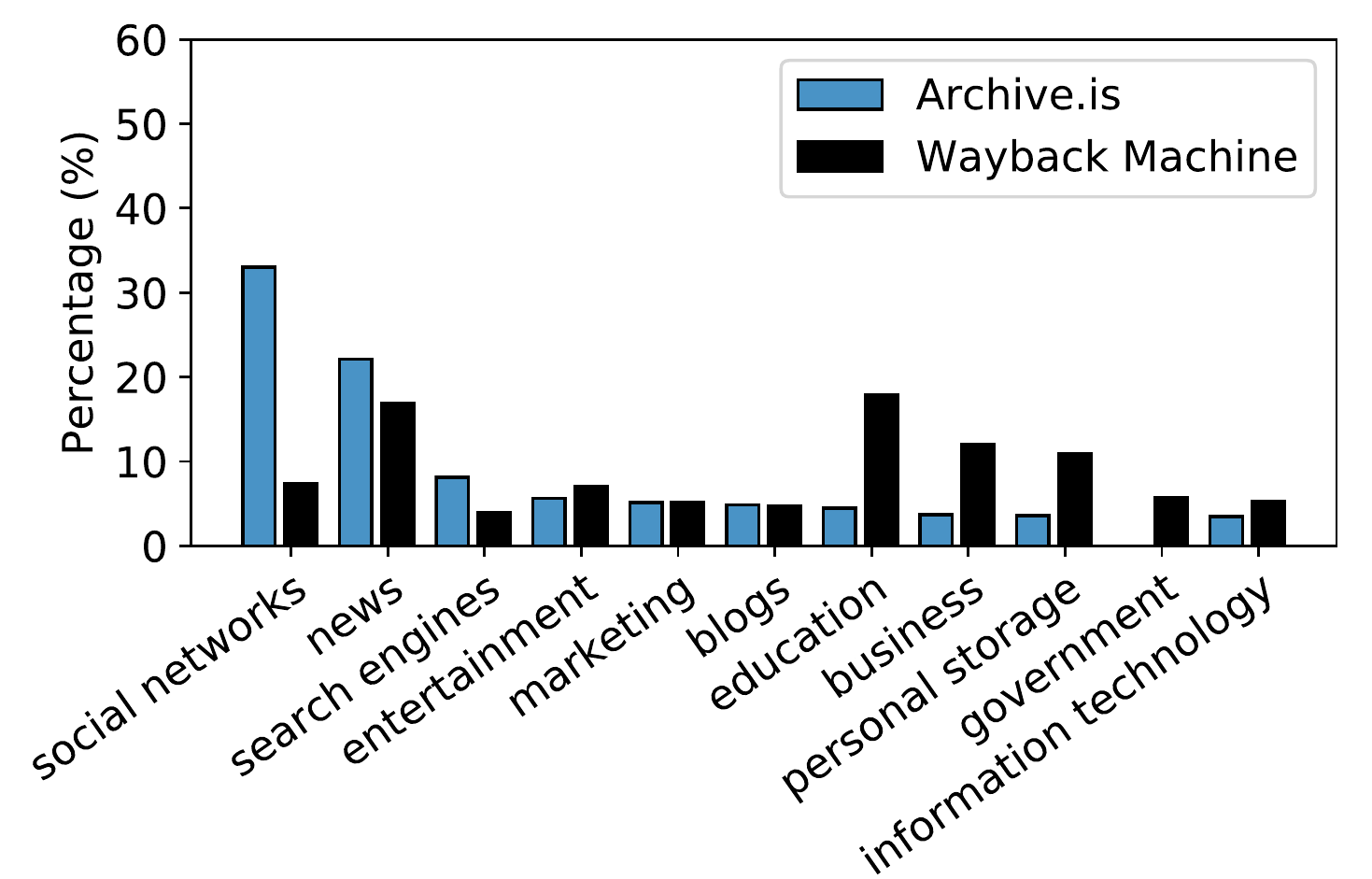}\label{subfig:bc_combined_twitter_virus}}\\
\subfigure[\dspol]{\includegraphics[width=0.495\columnwidth]{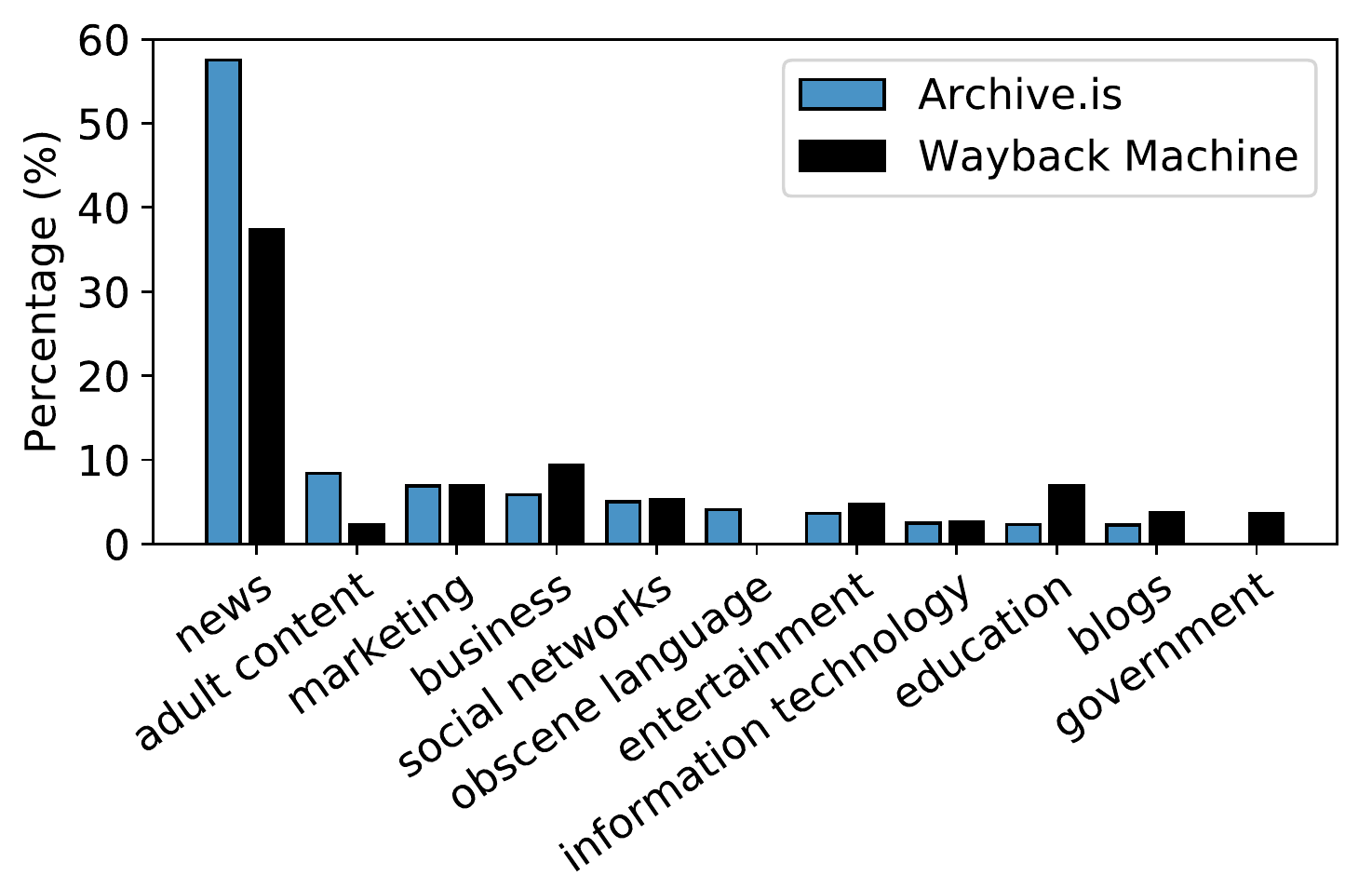}\label{subfig:bc_combined_4chan_virus}}
\subfigure[Gab]{\includegraphics[width=0.495\columnwidth]{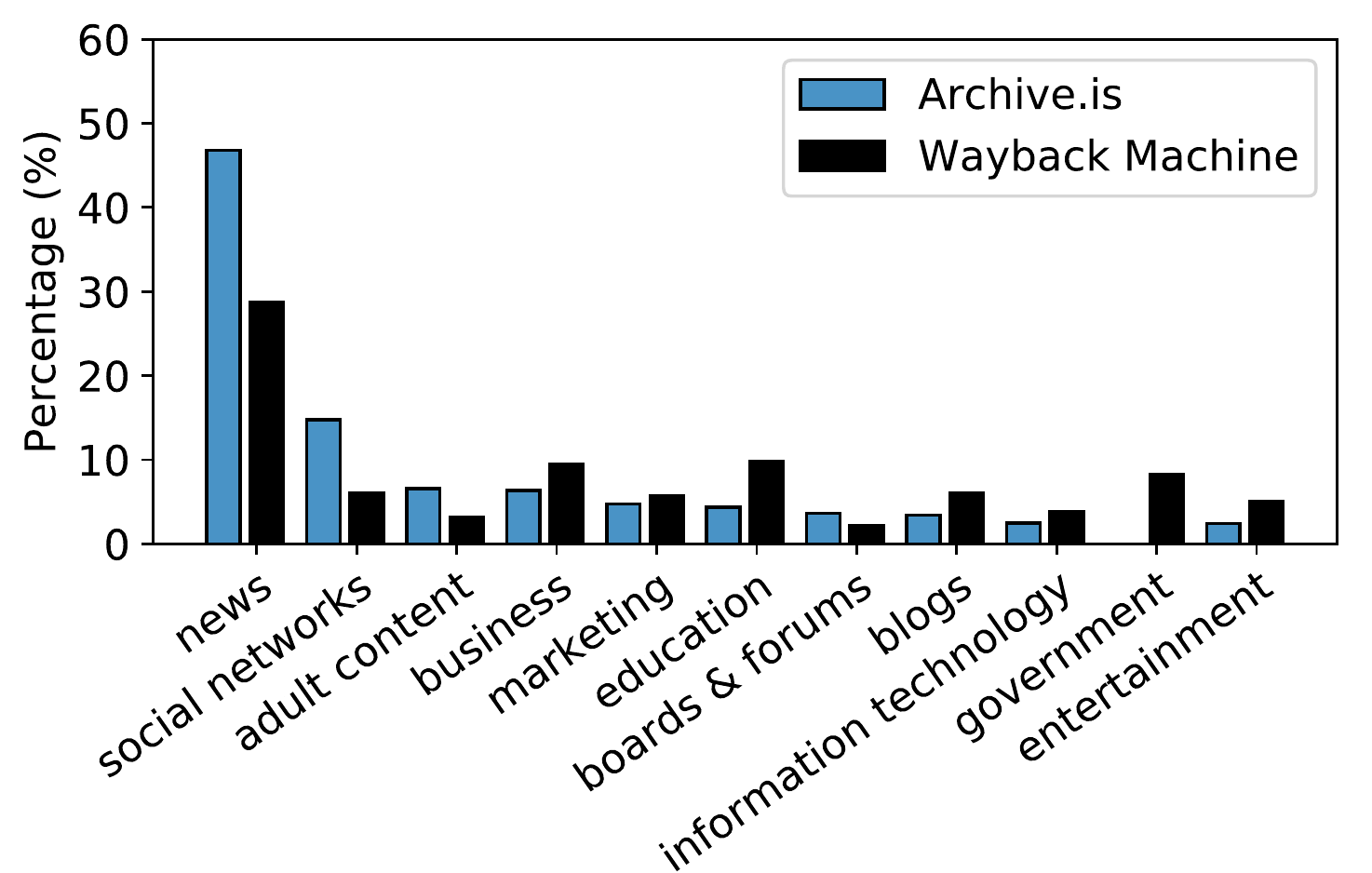}\label{subfig:bc_combined_gab_virus}}
\caption{Top domain categories for archive URLs appearing on the four social networks.}
\label{fig:top_categories}
\end{figure}

\descr{Social Networks.} Unlike the live feed dataset, we perform URL characterization for \textit{all} source URLs (aggregated by domain) found on Reddit, \dspol, Gab, and Twitter, again using the Virus Total API.
In Fig.~\ref{fig:top_categories}, we report the top categories and their corresponding percentages for both archiving services (specifically, the union of categories that appear in the top 10 categories for each service).
The Virus Total API is unable to provide a category for, on average, 1.5\% and 9\% of the archive.is and Wayback Machine 
URLs found on Reddit, \dspol, Gab, and Twitter, respectively.
Overall, both archiving services are often used to disseminate URLs from news sources, social networks, and marketing sites on all social networks.
However, there are interesting differences for the two archiving services:
Education and Government URLs appear as top categories for the Wayback Machine 
(see Fig.~\ref{subfig:bc_combined_twitter_virus},~\ref{subfig:bc_combined_4chan_virus}, and~\ref{subfig:bc_combined_gab_virus}), 
while sites that contain obscene language appear only for archive.is (see Fig.~\ref{subfig:bc_combined_4chan_virus}).
This suggests that the latter is used more extensively for ``questionable'' content.
Moreover, we observe that Adult Content is among the top categories for all social networks except Twitter,
while Gab and Reddit users often share archive URLs for domains related to Boards and Forums.
Also, on \dspol, archive.is is used to archive and disseminate pages with obscene language,
which is somewhat in line with previous observations~\cite{hine2017kek} showing that  
\dspol conversations often include hate speech. %

\subsection{Temporal Dynamics}
Next, we study, from a temporal point of view, how archive URLs are created and shared on social networks. 

\begin{figure}[t!]
\subfigure[Date]{\includegraphics[width=0.495\columnwidth]{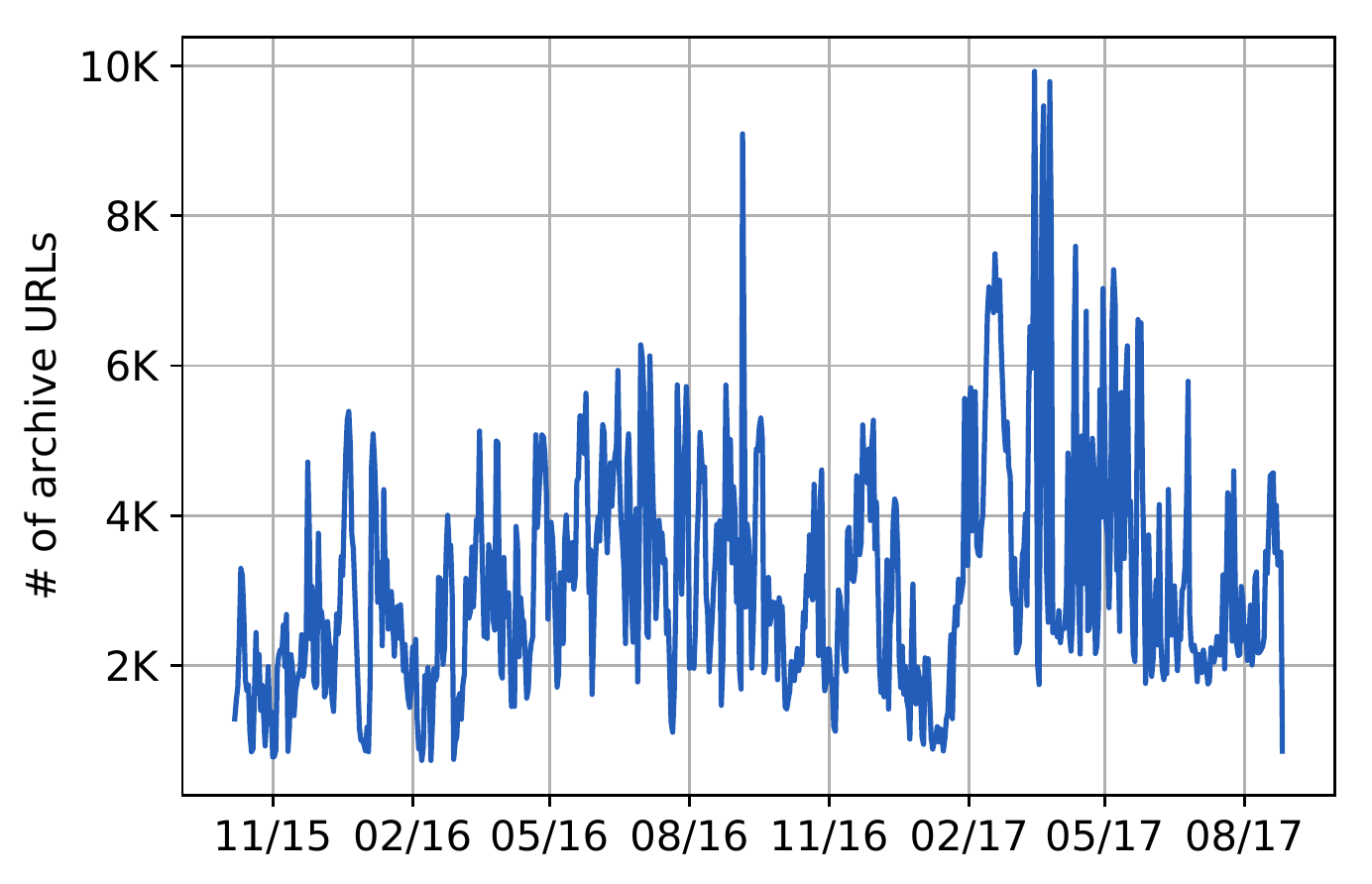}\label{subfig:counts_day}}
\subfigure[Hour of Day]{\includegraphics[width=0.495\columnwidth]{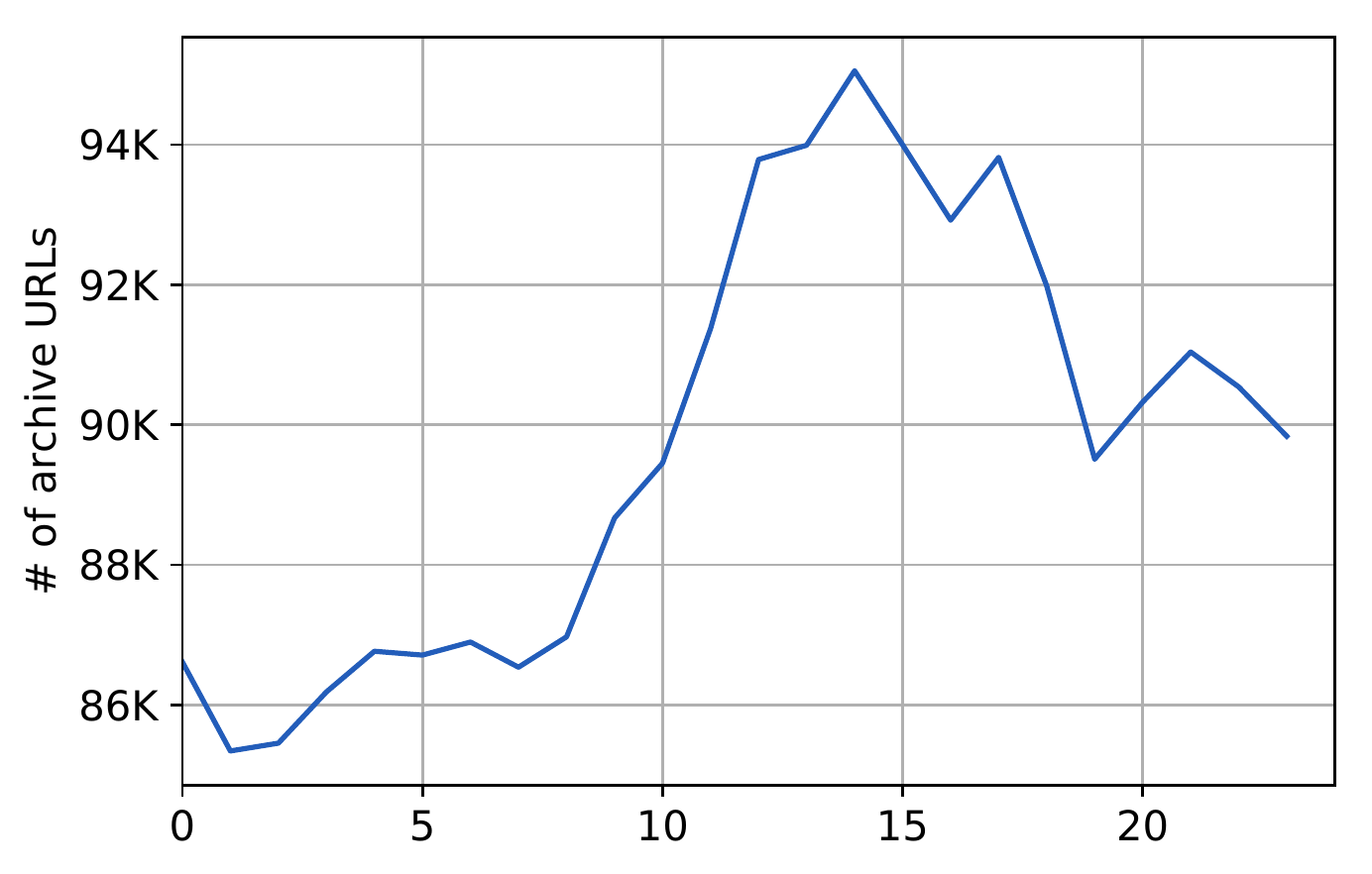}\label{subfig:counts_hour_day}}
\caption{Temporal analysis of the archive.is live feed dataset, reporting the number of URLs that are archived (a) each day and (b) based  on hour of day.}
\label{fig:temporal_analysis}
\end{figure}

\descr{Live Feed.}  In Fig.~\ref{fig:temporal_analysis}, we plot the day and hour of day of the creation of the archive.is URLs.
Each day, between 1K and 10K URLs are archived (Fig.~\ref{subfig:counts_day}), mostly between 11AM and 4PM UTC time, with a peak at 2PM (Fig.~\ref{subfig:counts_hour_day}), which seems to suggest that a great number of users may be located in Europe and the US. 
According to Alexa, the top country for archive.is is the US, with 37\% of the visitors. %

\begin{figure}[t]
\subfigure[archive.is]{\includegraphics[width=0.495\columnwidth]{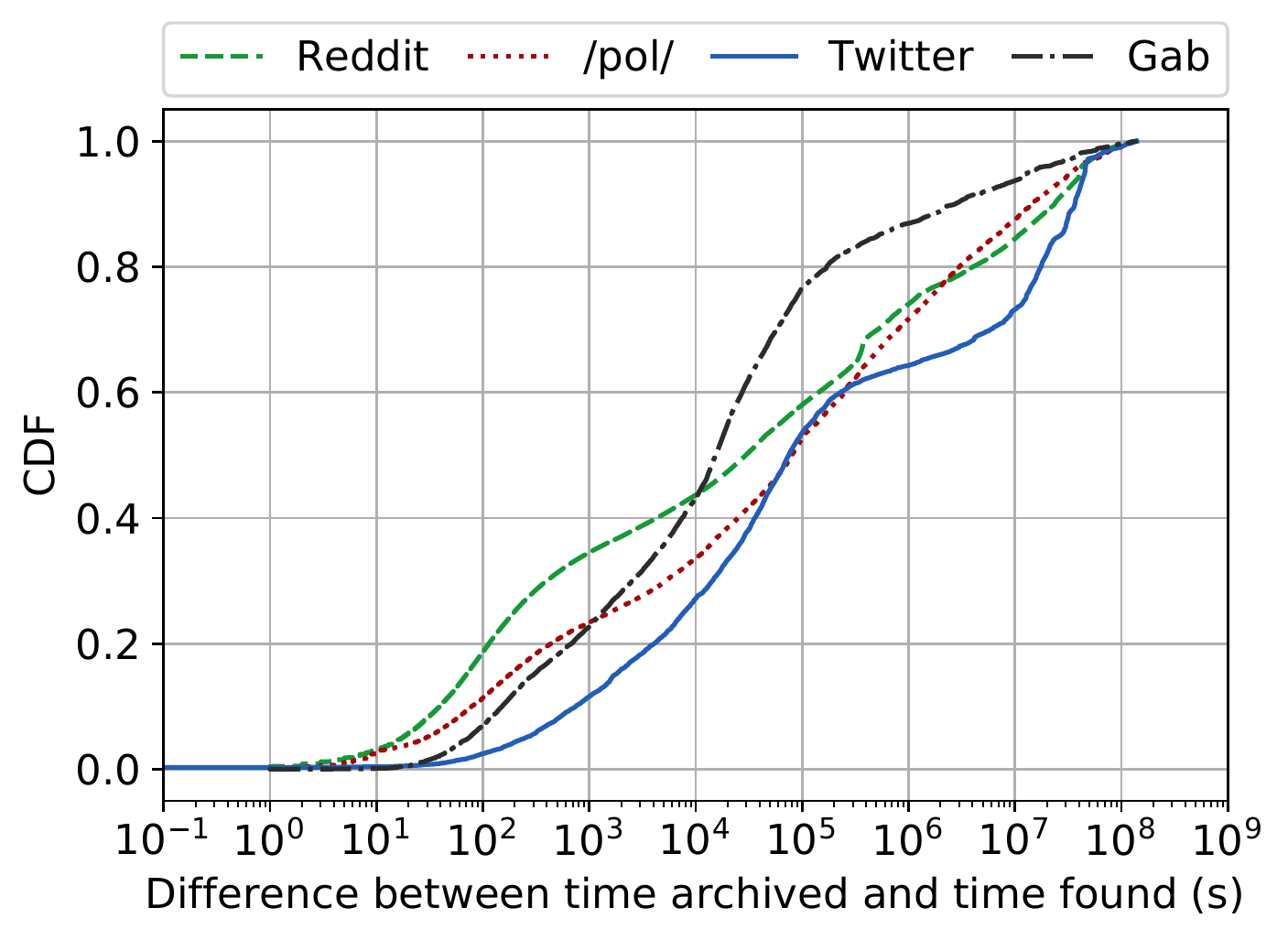}\label{subfig:cdf_time_difference_archive_is}}
\subfigure[Wayback Machine]{\includegraphics[width=0.495\columnwidth]{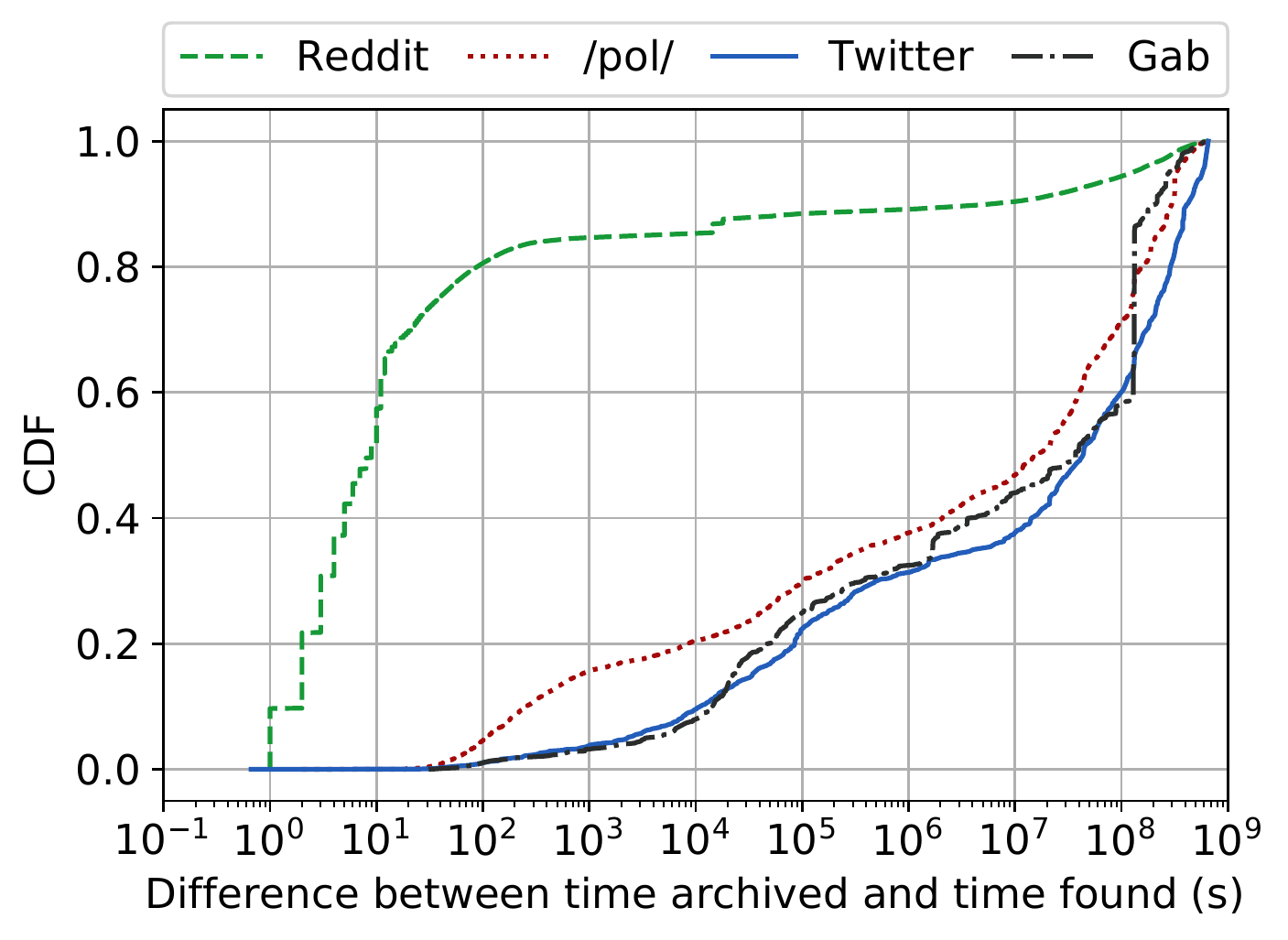}\label{subfig:cdf_time_difference_archive_org}}
\caption{CDF of the time difference between the archival time and the time appeared on each of the four social networks. }%
\label{fig:cdf_time_difference}
\end{figure}

\descr{Social Networks.} Next, we measure the time interval between the archiving of a URL and its appearance on one of the four social networks.
In Fig.~\ref{fig:cdf_time_difference}, we plot the CDF of these time intervals, finding that the interval between archiving and sharing times of a URL ranges from a few seconds (in which case, Reddit/4chan/Twitter/Gab users themselves might be creating the archive) to years.
Reddit is the ``fastest'' platform for Wayback Machine URLs, mainly because of bots that actively archive URLs (as we show later in the paper), while for archive.is it is Gab.

\begin{figure*}[t]
\center
\subfigure[Reddit]{\includegraphics[width=0.23\textwidth]{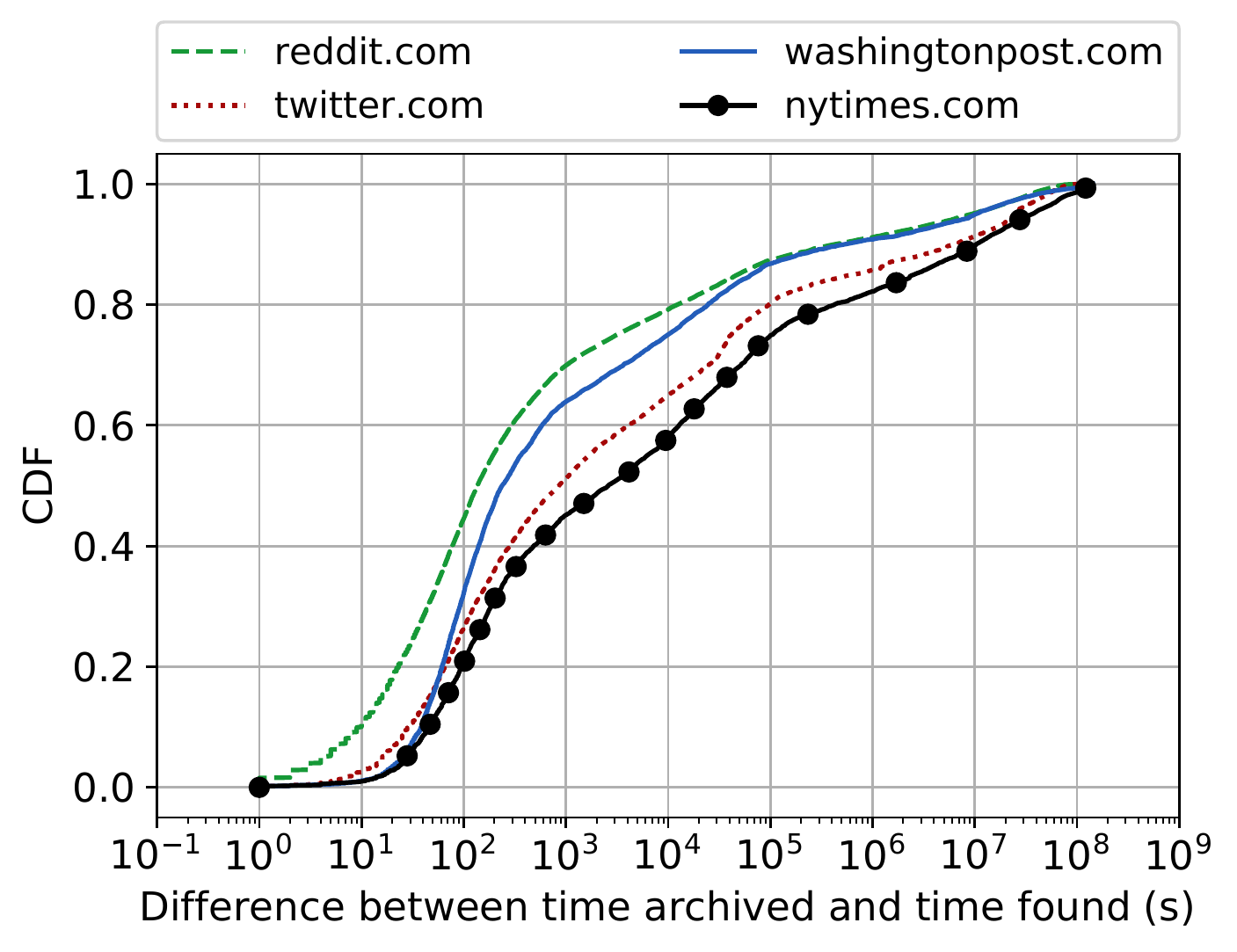}\label{subfig:cdf_time_difference_domain_archive_is_reddit}}
\subfigure[Twitter]{\includegraphics[width=0.23\textwidth]{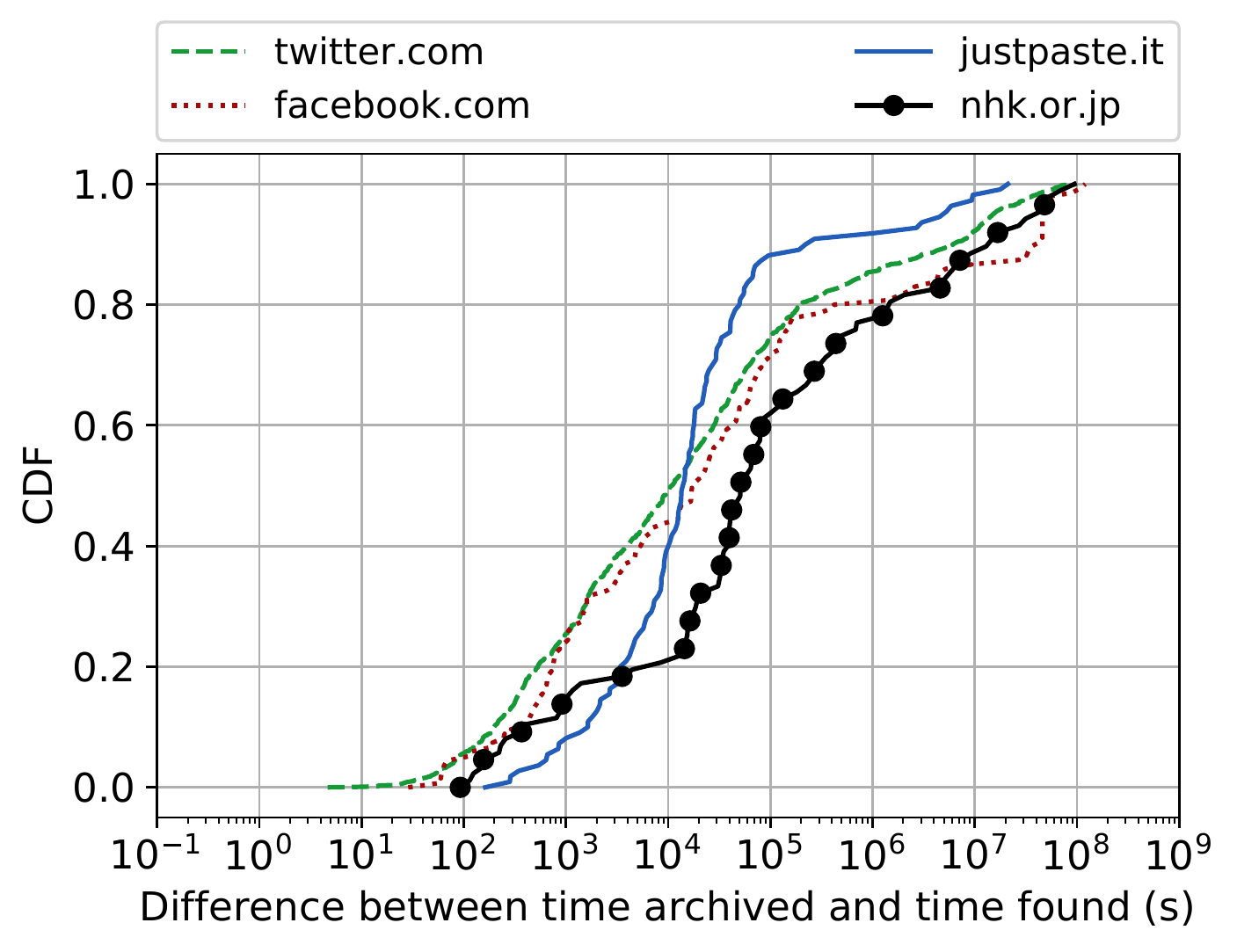}\label{subfig:cdf_time_difference_domain_archive_is_twitter}}
\subfigure[\dspol]{\includegraphics[width=0.23\textwidth]{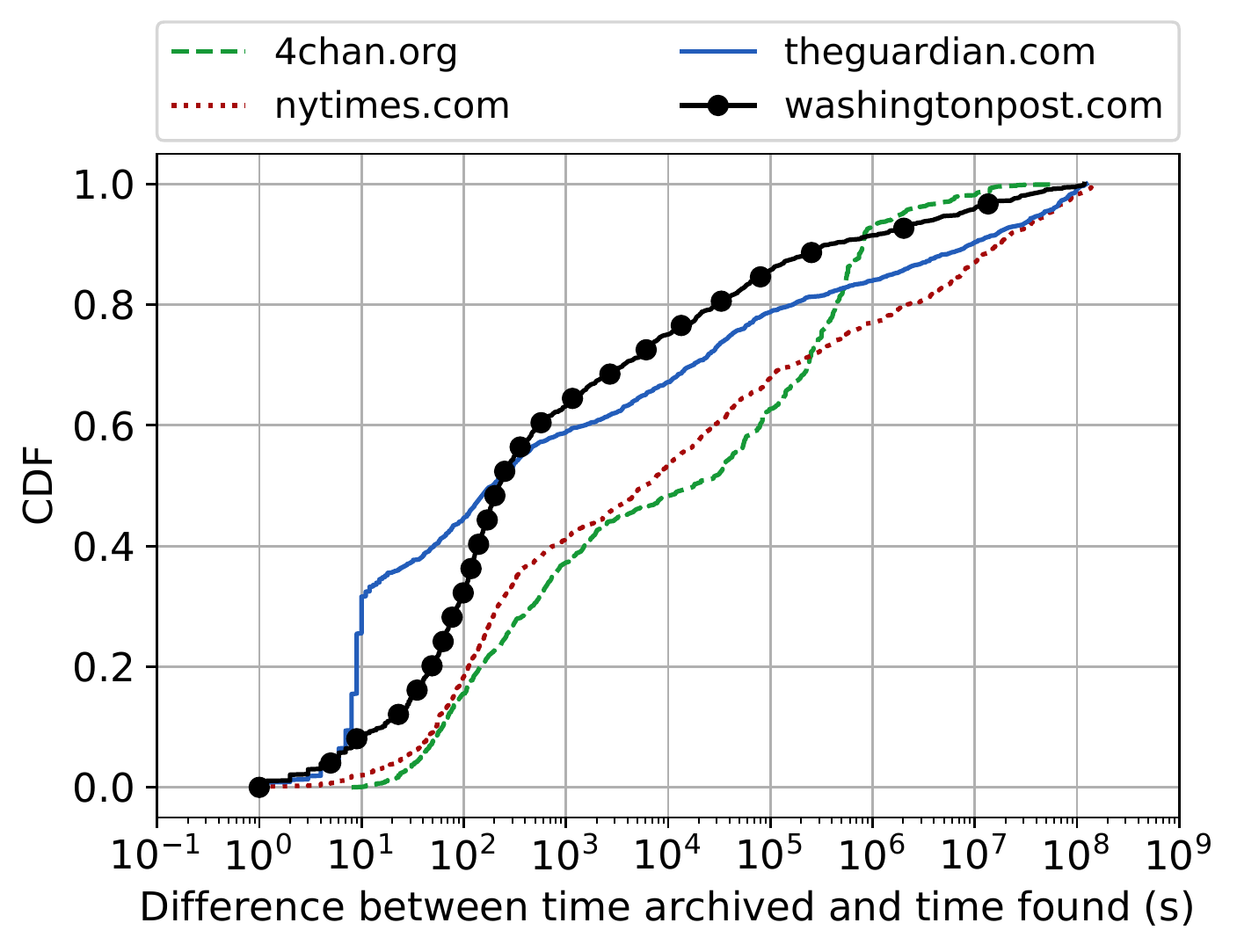}\label{subfig:cdf_time_difference_domain_archive_is_4chan}}
\subfigure[Gab]{\includegraphics[width=0.23\textwidth]{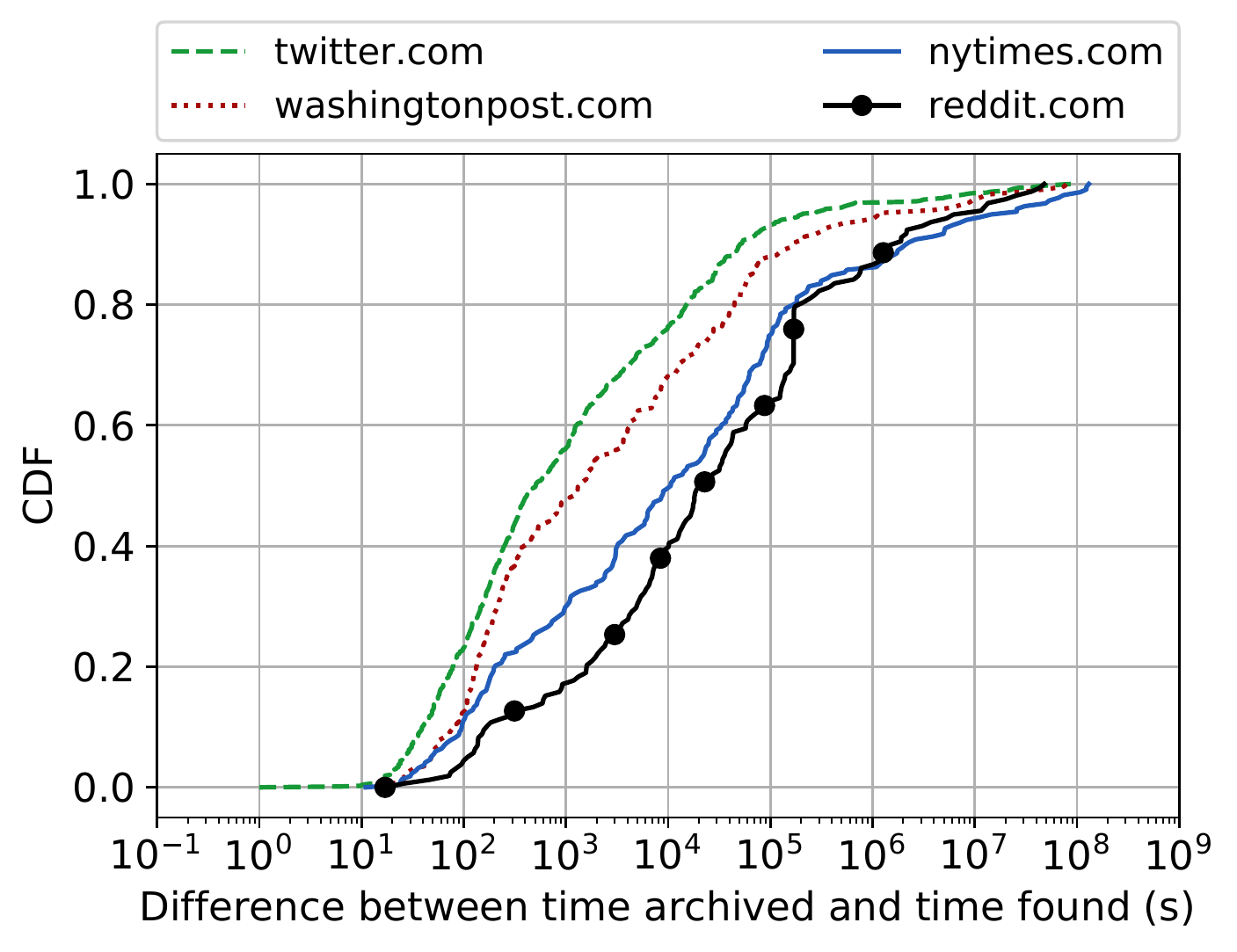}\label{subfig:cdf_time_difference_domain_archive_is_gab}}
\caption{CDF of the time difference between archival time on archive.is and appearance on social networks for the top four source domains.}
\label{fig:cdf_time_difference_domain_archive_is} 
\end{figure*}

\begin{figure*}[t]
\center
\subfigure[Reddit]{\includegraphics[width=0.23\textwidth]{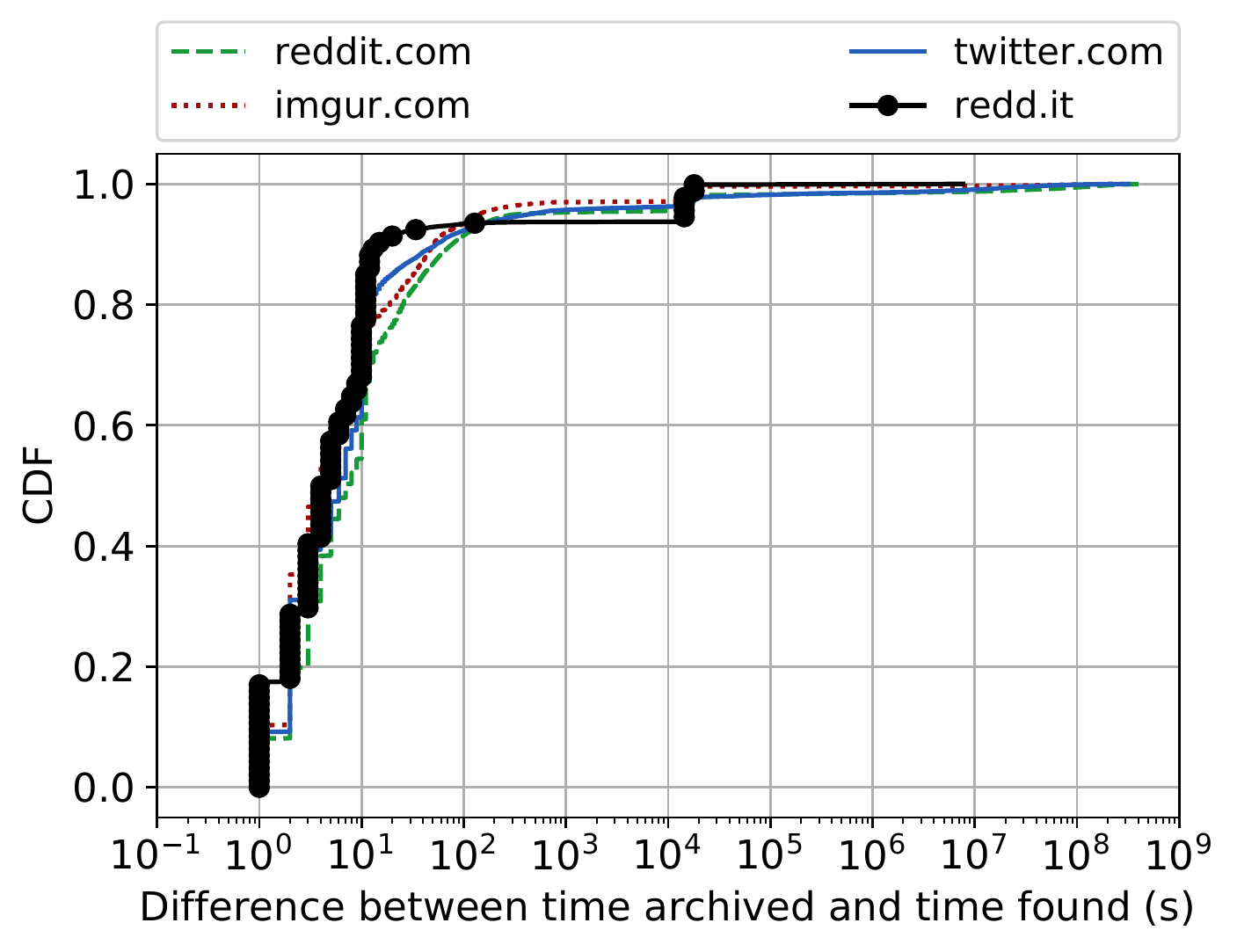}\label{subfig:cdf_time_difference_domain_archive_org_reddit}}
\subfigure[Twitter]{\includegraphics[width=0.23\textwidth]{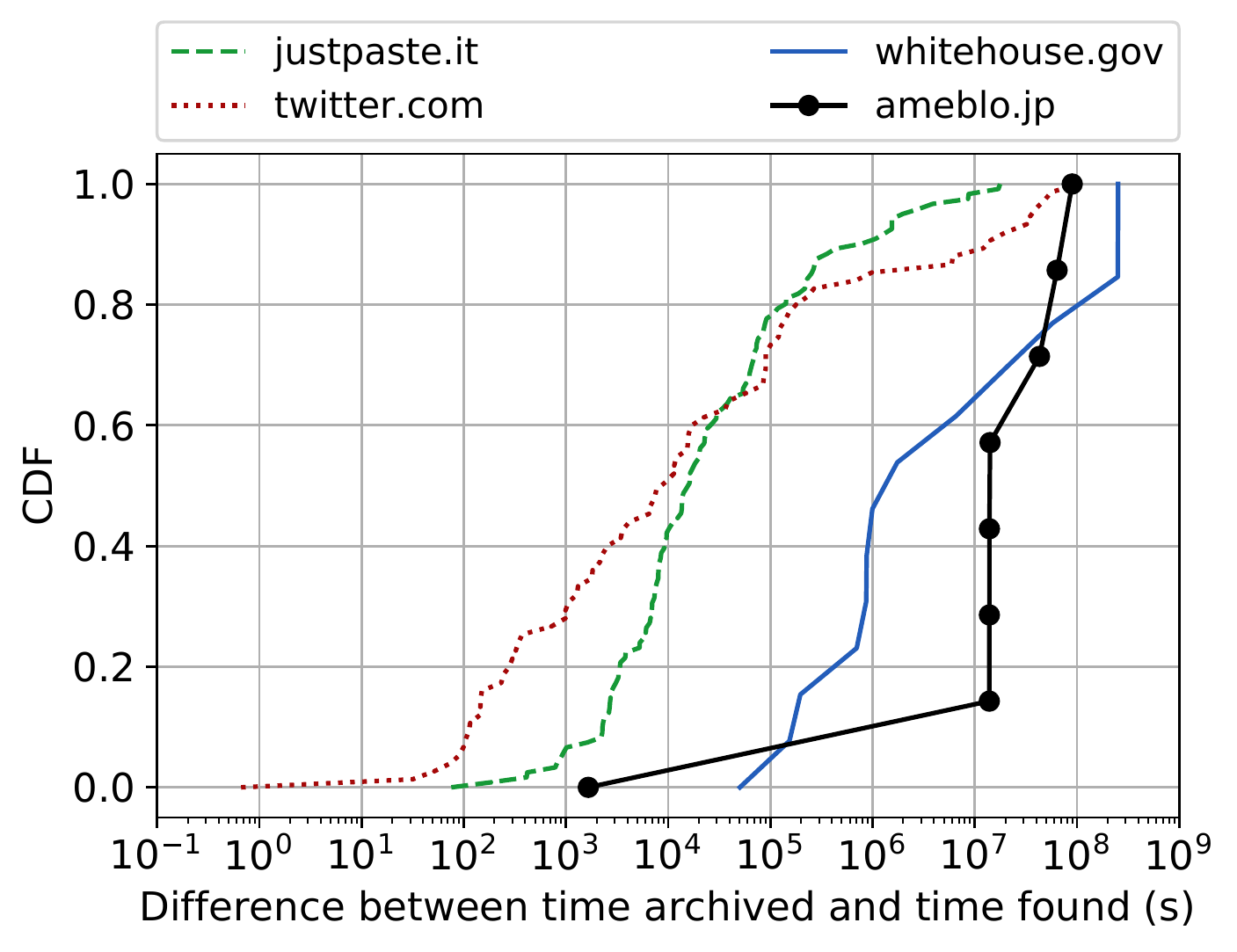}\label{subfig:cdf_time_difference_domain_archive_org_twitter}}
\subfigure[\dspol]{\includegraphics[width=0.23\textwidth]{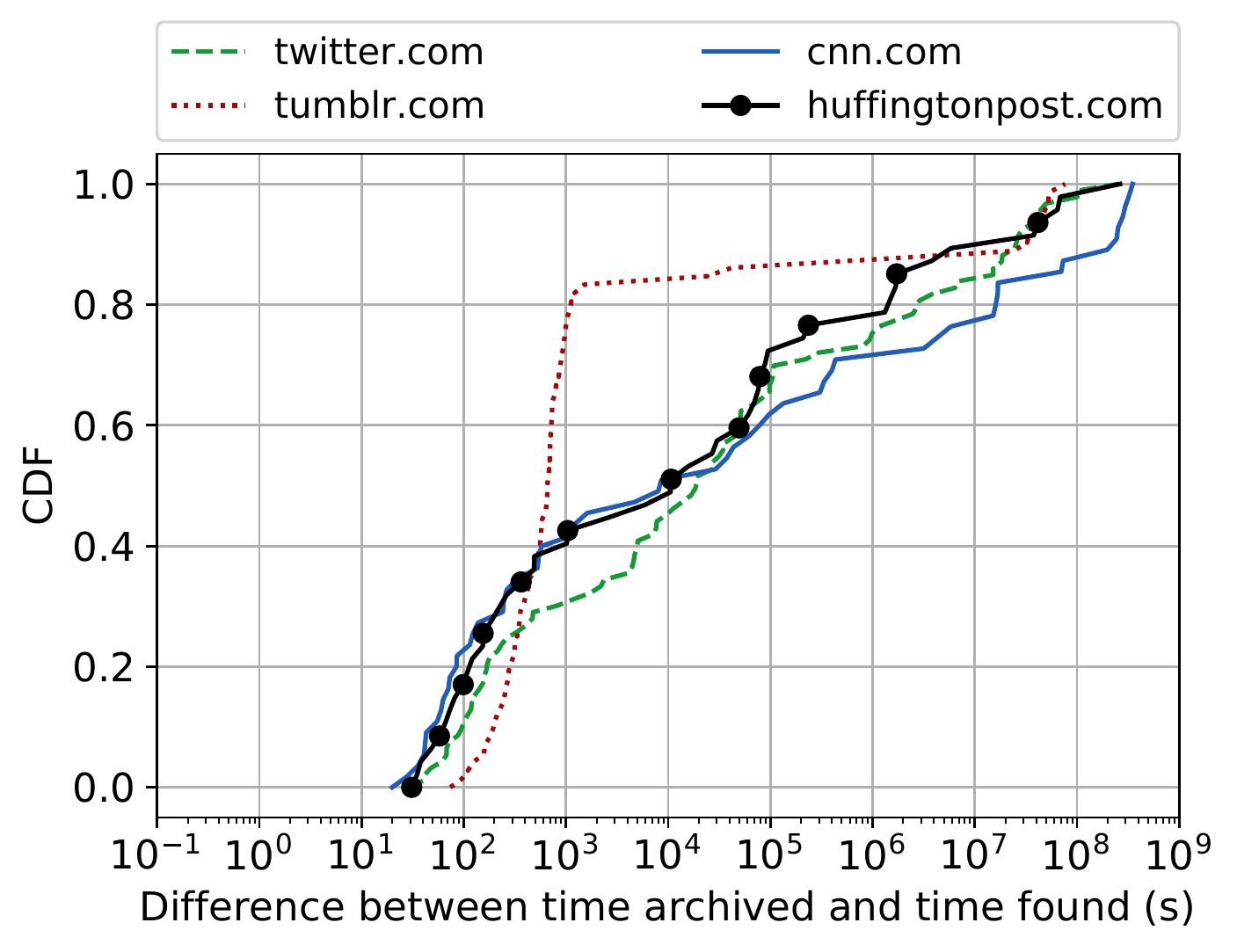}\label{subfig:cdf_time_difference_domain_archive_org_4chan}}
\subfigure[Gab]{\includegraphics[width=0.23\textwidth]{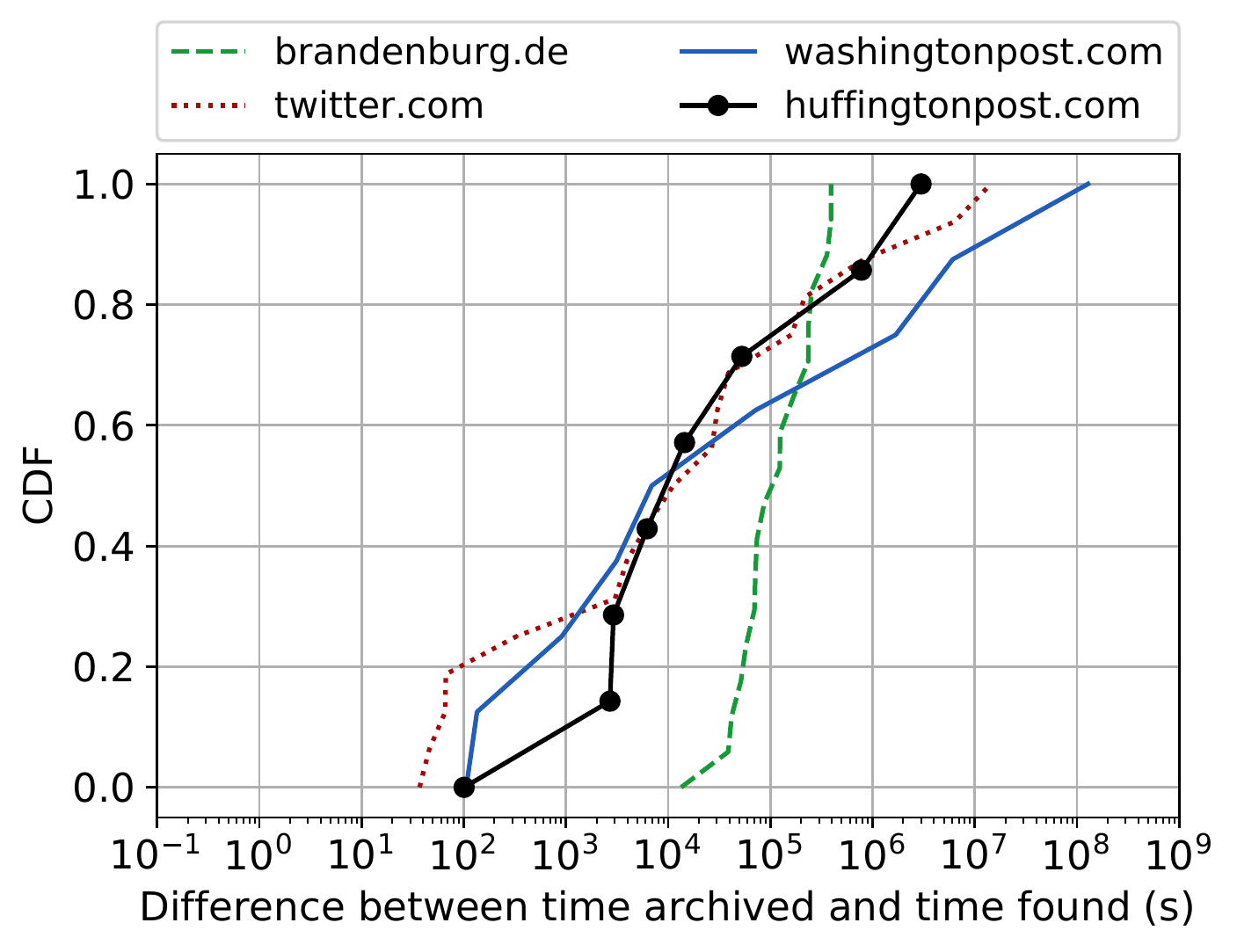}\label{subfig:cdf_time_difference_domain_archive_org_gab}}
\caption{\hspace{-0.115cm} CDF \hspace{-0.1cm} of \hspace{-0.1cm} the \hspace{-0.1cm} time \hspace{-0.1cm} difference \hspace{-0.1cm} between \hspace{-0.1cm} archival \hspace{-0.1cm} time \hspace{-0.1cm} on \hspace{-0.1cm} Wayback \hspace{-0.1cm} Machine \hspace{-0.1cm} and \hspace{-0.1cm} appearance \hspace{-0.1cm} on \hspace{-0.1cm} social \hspace{-0.1cm} networks \hspace{-0.1cm} for \hspace{-0.1cm} top \hspace{-0.1cm} four \hspace{-0.1cm} source \hspace{-0.1cm}  domains.}
\label{fig:cdf_time_difference_domain_archive_org} 
\end{figure*}

We also focus on the top source domains shared via archive URLs: 
Figs~\ref{fig:cdf_time_difference_domain_archive_is}--\ref{fig:cdf_time_difference_domain_archive_org} plot the CDF of the \emph{slack time} of the top four domains for archive.is and Wayback Machine URLs, respectively.
We define slack time as the time difference between the archival time on archiving services and the appearance on other social media.
On Reddit, the top domains archived via Wayback Machine follow very similar distributions, likely due to bots, while for 
archive.is URLs, distributions vary, with the fastest domain being \url{reddit.com} itself.
On Twitter, slack times vary for URLs archived via archive.is, with the fastest domain being Twitter and the slowest \url{nhk.org.jp}. 
The same applies for the Wayback Machine, with the fastest domain being Twitter and the slowest \url{ameblo.jp}.
We also find that, on \dspol, archive.is URLs pointing to 4chan are considerably slower, suggesting that users are more interested in archiving the URL for persistence rather than sharing the content within \dspol. Based on anecdotal observations, we believe users might be archiving threads with ``evidence'' for conspiracy theories/false narratives, and using them in the future to perpetuate mis/disinformation.
This is not the case for news sources like the Washington Post or Guardian, as \dspol users might be more focused on reducing Web traffic to the source domain instead (indeed we find users explicitly mentioning this when manually examining posts).
Finally, on Gab, the faster domain is Twitter, and Reddit the slowest.

\subsection{Original Content Availability}
We then assess the availability of the original archived content; this allow us to determine whether users are archiving URLs that are subsequently deleted. 
To this end, we make an HTTP request for each source URL in our datasets, on Oct 14--21, 2017 for the live feed dataset, on Oct 4--5, 2017 for Reddit, Twitter, \dspol datasets and on Jan 3, 2018 for Gab dataset.
We treat each URL as unavailable if we receive HTTP codes 404/410/451/5xx, or if the request times out.

\descr{Live Feed.} 
We find that 12\% of the source URLs corresponding to archive URLs on archive.is live feed are no longer available.
Domains with most unavailable content include \url{twitter.com} (6\%), \url{nhk.or.jp} (6\%), \url{googleusercontent.com} (3\%), \url{aaaaarg.fail} (3\%), and \url{4chan.org} (3\%).%

\descr{Social Networks.}
In Reddit, source URLs corresponding to both archive.is and Wayback Machine are still available to a large degree (93\% and 89\% of them, respectively). 
This can be explained by the fact that Reddit bots archive URLs without considering the content.
In \dspol, 82\% and 66\% of the original content is available for archive.is and Wayback Machine URLs, while on Gab it is 87\% and 48\%.
Percentages decrease further for Twitter, with 76\% and 49\% for archive.is and Wayback Machine URLs, respectively.

We also find that the top domains for which content is no longer available differ across platforms.
Except for Gab, the top unavailable domain are the social networks themselves: 10\%, 54\%, and 28\%, for Reddit, \dspol, and Twitter, respectively.
URLs from cache servers (i.e., \url{googleusercontent.com}) and Twitter are also frequently unavailable;  9\% and 10\% in Reddit,
5\% and 4\% in \dspol,  8\% and 28\% in Twitter, and 12\% and 19\% in Gab, for \url{googleusercontent.com} and Twitter, respectively.
We also note the presence of unavailable 8ch.net URLs (another ephemeral imageboard) with 5\% and 4\% on \dspol and Gab, respectively.

\subsection{Take-Aways}
Overall, we find that archiving services play an important role in the information ecosystem, as they are used to preserve news sources as well 
as ephemeral or controversial content. Also, users on fringe communities such as \dspol and Gab favor less popular Web archiving services 
like archive.is to archive and disseminate Web pages.
This prompts questions as to \emph{why} less popular, and seemingly less durable, archiving services are favored by more controversial 
communities like \dspol and Gab. Although this would be out of the scope of this work, we do find one potential answer in that these 
communities also use archiving services to bypass platform-specific censorship policies.

We also observe that temporal dynamics of how archive URLs are shared on social networks differ according to their content: for instance, on \dspol, content 
from news sources has a considerably larger time lag between first appearing on the platform and archival compared to 4chan threads.
Lastly, a non-negligible percentage of archived content is no longer available at the source; in particular, a substantial percentage of posts from social networks 
like Twitter are eventually deleted from the platform, yet remain stored in the archives.

\section{Social-Network-based Analysis} \label{sec:social_network_analysis}
In this section, we present a social-network-specific analysis by considering the fundamental differences of each platform.
We analyze the users involved in the dissemination of archive URLs as well as the content that is shared along with those URLs.
Lastly, we discuss a case study of ad revenue deprivation on Reddit.
Due to space limitations, we exclude Twitter from our analysis, given the relatively 
small footprint of archive URLs on that platform (see~Table~\ref{tab:dataset}).

\subsection{User Base}

\begin{figure}[t]
\subfigure[Reddit]{\includegraphics[width=0.48\columnwidth]{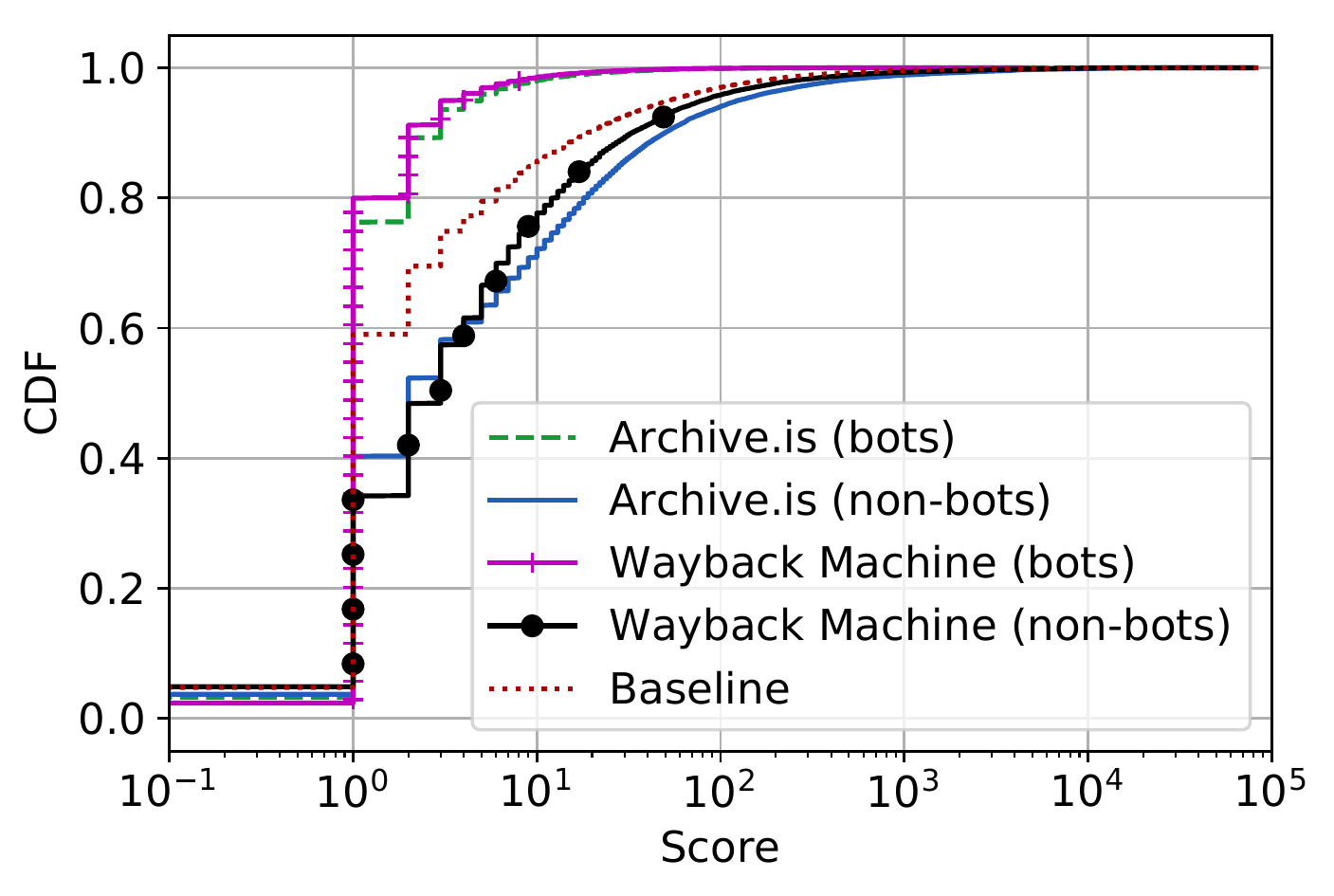}\label{subfig:cdf_scores_reddit_bots}}
\subfigure[Gab]{\includegraphics[width=0.48\columnwidth]{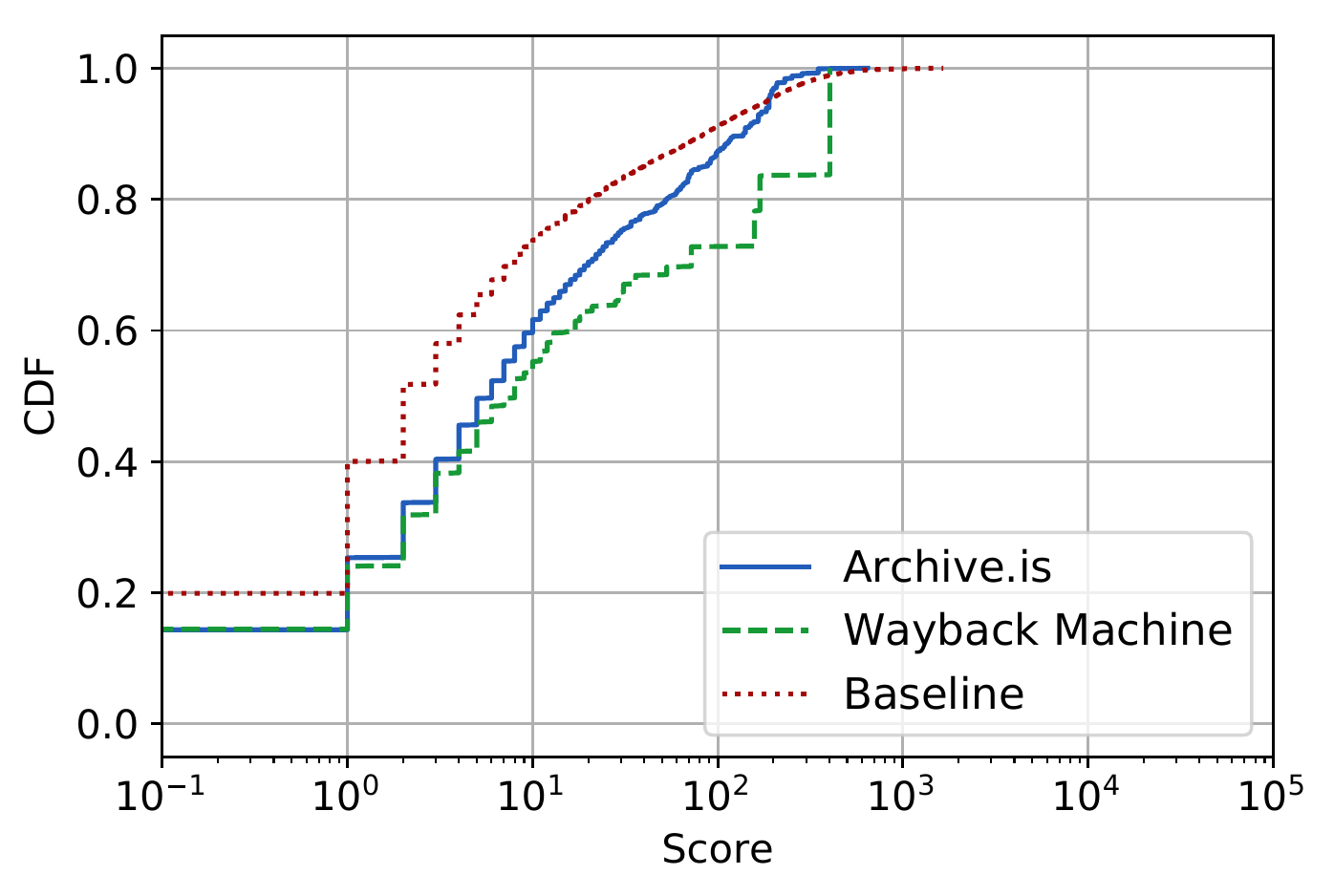}\label{subfig:cdf_scores_gab}}
\caption{CDF of the scores of posts that include archive.is and Wayback Machine URLs.}
\label{fig:score_analysis}
\end{figure}

\noindent\textbf{Reddit.} Our analysis shows that archiving services are extensively used by Reddit bots.
In fact, 31\% of all archive.is URLs and 82\% of Wayback Machine URLs in our Reddit dataset are posted by a specific bot, 
namely, SnapshillBot (which is used by subreddit moderators to preserve ``drama-related'' happenings discussed earlier or just as 
a subreddit specific policy to preserve \emph{every} submission).
Other bots include AutoModerator, 2016VoteBot, yankbot, and autotldr.
We also attempt to quantify the percentage of archive URLs posted from bots, assuming that, if a username includes ``bot'' or ``auto,'' it is likely a bot.
This is a reasonable strategy since Reddit bots are extensively used for moderation purposes, and do not usually try to obfuscate the fact that they are 
bots.\footnote{This is somewhat evident from the list of Reddit bots available at \url{https://www.reddit.com/r/autowikibot/wiki/redditbots}}
Using this heuristic, we find that bots are responsible for disseminating 44\% of all the archive.is and 85\% of all the Wayback Machine URLs 
that appear on Reddit between Jul 1, 2016 and Aug 31, 2017.
We also use the score of each Reddit post to get an intuition of users' appreciation for posts that include archive URLs.
In Fig.~\ref{subfig:cdf_scores_reddit_bots}, we plot the CDF of the scores of posts with archive.is and Wayback Machine URLs, as 
well as all posts that contain URLs as a baseline, differentiating between bots and non-bots.
For both archiving services, posts by bots have a substantially smaller score: 
80\% of them have score of at most one, as opposed to 37\% for non-bots and 59\% of the baseline.

\begin{table}[t]
\centering
\footnotesize
\resizebox{0.9\columnwidth}{!}{%
\begin{tabular}{lr|lr}
\toprule
\textbf{Subreddit (archive.is)} & \multicolumn{1}{r}{\bf (\%)} & \textbf{Subreddit (Wayback)} & \textbf{(\%)} \\ \midrule
The\_Donald                  & {24.48\%} & EnoughTrumpSpam  & 31.82\%     \\
KotakuInAction                  & {15.83\%}  & MGTOW & 7.38\%      \\
EnoughTrumpSpam                   & {12.06\%}  & SnapshillBotEx   &  7.19\%       \\
MGTOW              & {3.48\%}  & undelete   & 5.90\%        \\
undelete                & {2.74\%}  & SubredditDrama  & 5.50\%          \\
SubredditDrama                & {2.61\%}  & Drama   & 5.03\%          \\
Drama                    & {2.33\%}  & Gamingcirclejerk  & 3.47\%          \\
Gamingcirclejerk                  & {1.57\%}  & ShitAmericansSay   & 1.63\%          \\
conspiracy                  & {1.44\%}  & TopMindsOfReddit   & 1.51\%          \\
MensRights                  & {1.12\%}  & TheBluePill   & 1.25\%          \\
savedyouaclick               & {1.00\%}  & Buttcoin\_1000   & 1.15\%           \\
politics                  & {0.98\%}  &  AgainstHateSubreddits & 1.06\%            \\
DerekSmart                  & {0.76\%}  & subredditcancer   & 0.99\%            \\
ShitAmericansSay                    & {0.75\%}  & The\_Donald   & 0.95\%            \\
PoliticsAll                  & {0.72\%}  & badeconomics  & 0.75\%            \\ \bottomrule
\end{tabular}
}
\caption{Top 15 subreddits sharing archive.is and Wayback Machine URLs.}
\label{tbl:top_subreddits}
\end{table}

\descr{Reddit Sub-Communities.} We then study how specific subreddits share URLs from archiving services. 
In Table~\ref{tbl:top_subreddits}, we report the top subreddits that share the most archive URLs from archive.is and the Wayback Machine.
Among these, we find a variety of subreddits ranging from politics (e.g., EnoughTrumpSpam, The\_Donald) to gaming (e.g., Gamingcirclejerk) and 
``drama-related'' communities (e.g., SubredditDrama and Drama). 
Several subreddits prefer to use archive.is rather than the Wayback Machine, e.g., KotakuInAction, which historically covers 
the GamerGate controversy~\cite{chatzakou2017hate}, The\_Donald, which discusses politics with a focus on Donald Trump, and Conspiracy, 
which focuses on various conspiracy theories. 

\descr{Gab.} On Gab, each post has a score that determines the popularity of the content. In Fig.~\ref{subfig:cdf_scores_gab}, we report the CDF of 
the scores in posts that contain archive.is and Wayback Machine URLs, between August 2016 and August 2017.
Once again, we also include a baseline, which is the scores for all the posts with URLs.
We find that posts with Wayback Machine URLs have higher scores than those with archive.is URLs, and the baseline.
Specifically, the mean score for Wayback Machine is 90, while for archive.is and the baseline the mean score is 35 and 30, respectively.
This trend mirrors the one observed on Reddit for posts not authored by bots.

\descr{/pol/.} As mentioned earlier, 4chan is an anonymous imageboard, which prevents us from performing user-level analysis.
However, we can use the flag attribute to provide a country-level estimation. %
The top country sharing archive URLs is the USA, which is in line with previous characterizations of the board~\cite{hine2017kek}. 
We also find a substantial percentage of ``troll'' flags: 9\% and 5\% for archive.is and Wayback Machine, respectively (see Sec. Background for a description of ``troll'' flags).
This is somewhat surprising, since troll flags were re-introduced to \dspol on June 13, 2017, thus they were only available for about 
3 months of our 14-month dataset.

\subsection{Content Analysis}
\begin{figure}[t]
\centering
\includegraphics[width=0.55\columnwidth]{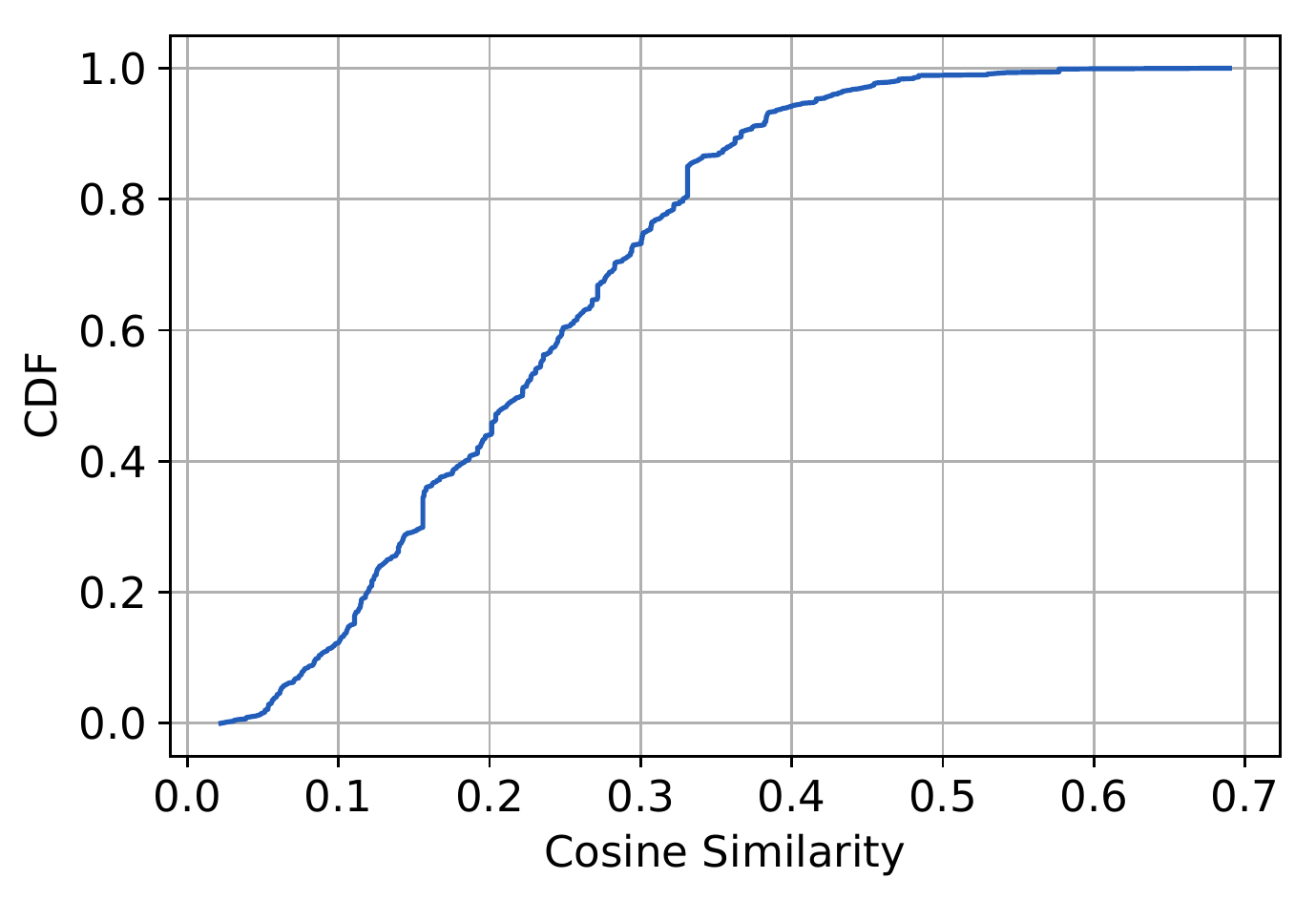}
\caption{CDF of cosine similarity of words obtained from LDA topics on Reddit and \dspol threads.}
\label{fig:cdf_lda_similarity}
\end{figure}

Next, we focus on the content that gets shared along with archive URLs on social platforms. We aim
to evaluate if users share the same information for a given archive URL on multiple platforms.
We do so using Latent Dirichlet Allocation (LDA).
Due to space limitations, we only study Reddit and \dspol.
Before running LDA, we exclude \dspol and Reddit threads that contain less than 100 posts, 
and select only threads that have archive URLs appearing in {\em both} Reddit and \dspol datasets;
there are 425 such threads on \dspol and 299 on Reddit.
Next, we run LDA on all the posts within these threads and extract terms for 10 topics per thread.
In Fig.~\ref{fig:cdf_lda_similarity}, we plot the CDF of the cosine similarities on the terms extracted from LDA topics
on the two platforms when sharing the same archive URLs.
We observe that 80\% of the terms have similarity under 0.3, which is expected given the fact that the two communities discuss topics in a 
different way. By manually observing terms with high similarity scores, we find that a number of them relate 
to well-known conspiracy theories, like the Seth Rich murder~\cite{seth_rich} or Pizzagate~\cite{bbc_4chan_pizzagate}, as well as general 
discussions around politics (e.g., tensions between North Korea and the USA).
Once again, this highlights that archiving services are used to preserve content related to controversial 
stories and conspiracy theories.  A more detailed analysis is deferred to the extended version of the paper.
 
\subsection{Ad Revenue Deprivation}

During our experiments, we find evidence that at least one Reddit bot, AutoModerator, is used to 
remove links to unwanted domains and nudge users to share archive.is instead. In particular, it posts:
``\emph{Your submission was removed because it is from \url{cnn.com}, which has been identified as a severely anti-Trump domain. 
Please submit a cached link or screenshot when submitting content from this domain. We recommend using \url{www.archive.is} for this purpose.}'' 

This kind of notification appears in five different subreddits that discuss mainly politics and news, specifically, 
The\_Donald, Mr\_Trump, TheNewRight, Vote\_Trump, and Republicans.
In particular, in The\_Donald, there are 13K such comments. AutoModerator blocks URLs from 23 news sources likely to be considered as 
anti-Trump by that community. In Table~\ref{tbl:reddit_deleted_domains} we report the number of submissions 
deleted for each of the sources, along with the percentage over {\em all} submissions that include that source.
Mainstream news outlets like Washington Post and CNN are the top domains that get removed from The\_Donald (3.8K and 3.3K submissions, respectively), 
and this happens slightly less than half the times (44\% and 39\% of the submissions, respectively).
Interestingly, only URLs posted via the URL submission field are censored by AutoModerator, 
but not URLs that are inserted as part of the title field.

\begin{table}[t]
\centering
\footnotesize
\resizebox{\columnwidth}{!}{%
\setlength{\tabcolsep}{0.25em} %
\begin{tabular}{lrr|lrr}
\toprule
\textbf{News Source} & \textbf{Count} & \multicolumn{1}{r}{(\%)} &\textbf{News Source} & \textbf{Count} & \textbf{ (\%)} \\ \midrule
washingtonpost.com          &      3,814  & {44.13\%} & change.org & 96    &  7.52\%\\
cnn.com       &         3,354  & {39.39\%}  & huffpost.com  &  62     & 13.39\%\\
nydailynews.com           &     1,070   & {46.32\%}  & fusion.net   &  60     & 44.77\%\\
huffingtonpost.com            &  978  & {43.77\%}  & cnn.it   & 58    & 44.61\% \\
nationalreview.com &            774   & {45.58\%}  & alternet.org  & 26   & 20.01\%      \\
theblaze.com               &  704  & {46.74\%}  & infostormer.com   & 16    &   27.11\%  \\
buzzfeed.com                   & 588 & {45.97\%}  & dailynewsbin.com   & 4         & 26.67\% \\
salon.com            &      373  & {44.88\%}  & todayvibes.com   & 4  &     7.27\%  \\
vice.com                  & 372 & {45.14\%}  & usanewsbets.ga   & 4       &  10.52\%\\
vox.com               & 323  & {45.23\%}  & fullycucked.com   & 1  & 1.78\%       \\
weeklystandard.com        &      253  & {46.25\%}  & northcrane.com   & 1         & 0.13\%\\
politifact.com              &    185 & {33.09\%}  &    &     &        \\ \bottomrule
\end{tabular}
}
\caption{Number and percentage of submissions deleted from The\_Donald with links to different news sources.}
\label{tbl:reddit_deleted_domains}
\end{table}

We attempt to estimate possible ad revenue deprivation due to the practice of forcing users to share archive URLs instead of source URLs on Reddit.
We do so by providing a conservative approximation of the ad revenue loss. Since we do not have knowledge of how many times a particular URL is 
clicked, we use the up- and down-votes of a post. That is, we assume that when a user up-votes or down-votes a post, he also clicks on the URL included on the post.
This constitutes a best-effort technique as prior work shows that a substantial portion of users on Reddit do not vote~\cite{gilbert2013underprovision}, while, at 
the same time, users that do vote do not necessarily read or click on the articles~\cite{glenski2017consumers}.
That said, this approach is reasonably conservative considering the influence that Reddit has with respect to news 
dissemination~\cite{zannettou2017web}.

We then calculate the potential revenue loss using only ad impressions, i.e., we conservatively estimate the revenue generated when a user 
visits the website without taking into account any potential further action (e.g., clicking on the actual ad).
To this end, we use an average Cost per 1,000 impressions (CPM) of \textdollar 24.74,
as reported by Statista\footnote{\url{https://www.statista.com/statistics/308015}}, 
while we assume an average of 3 ads per page~\cite{barford2014adscape}.
In other words, we calculate the monthly revenue loss, for each domain, based on the average CPM value as well as 
the conservative estimate of the visits using the up- and down-votes.
Overall, replacing URLs with archive URLs, as done, e.g., by the AutoModerator bot, yields an estimate of 
\textdollar 30K per month in revenue loss (for the top 20 domains in terms of views). This is detailed in Table~\ref{tbl:revenue_losses}, 
where we break down the estimate for each of the top 20 revenue-deprived domains.

On a purely pragmatic level, consider that our conservative estimate of ad revenue deprivation is around \textdollar 70K per year for the Washington Post alone.
Although a more detailed impact analysis is out of the scope of this work, we suspect that even \textdollar 70K could have a real world effect, 
e.g., on intern budgets or even early career hires.

\begin{table}[]
\centering
\resizebox{\columnwidth}{!}{
\setlength{\tabcolsep}{0.25em} %
\begin{tabular}{lrr|lrr}
\toprule
\textbf{ Domain} & \textbf{ Visits} & \multicolumn{1}{r}{\bf Loss (\$)} & \textbf{Domain} & \textbf{Visits} & \textbf{Loss (\$)} \\ \midrule
washingtonpost.com     & 79,880                  & 5,928                              & wsj.com                & 11,389                  &   845                             \\
cnn.com                & 70,483                  & 5,231                              & breitbart.com          & 11,357                  & 842                               \\
nytimes.com            & 46,442                  & 3,446                              & bbc.com                & 10,708                  & 794                               \\
huffingtonpost.com     & 27,125                  & 2,013                              & salon.com              & 10,364                  & 769                               \\
thehill.com            & 18,643                  & 1,383                               & buzzfeed.com           & 10,359                  & 768                               \\
theguardian.com        & 16,376                  & 1,215                               & foxnews.com            & 9,638                   & 715                               \\
politico.com           & 15,774                  &       1,170                         & yahoo.com              & 9,497                   & 704                              \\
dailymail.co.uk        & 14,442                  & 1,071                               & latimes.com            & 9,277                   & 688                               \\
dailycaller.com        & 12,735                  & 945                               & vox.com                & 8,976                   & 667                               \\
google.com             & 11,576                  & 859                               & washingtontimes.com    & 8,862                   & 657                               \\ \bottomrule
\end{tabular}
}
\caption{Top 20 domains with the largest ad revenue losses because of the use of archiving services on Reddit. 
We report an estimate of the average monthly visits from Reddit and the monthly ad loss. }
\label{tbl:revenue_losses}
\end{table}

\subsection{Take-Aways}
In summary, our social-network-specific analysis shows, among other things, that moderation bots on Reddit proactively 
leverage Web archiving services to ensure that content shared on their community persists.
In particular, we find that 44\% and 85\% of archive.is and Wayback Machine URLs are shared by Reddit moderation bots, respectively.
Also, Web archiving services are extensively used for the archival and dissemination of content related to conspiracy theories 
(e.g., Pizzagate) as well as other world events related to politics (e.g., tensions between North Korea and the USA), 
thus suggesting that these services play an important role in the (false) information ecosystem and need to be taken into 
account when designing systems to detect and contain the cascade of misinformation on the Web.
Finally, we find evidence that moderators from specific subreddits force users to misuse Web archiving services 
so as to ideologically target certain news sources by depriving them of traffic and potential ad revenues. 
We also provide a best-effort conservative estimate of ad revenue loss of popular news sources showing that they can lose up to \textdollar 70K per year.

\section{Discussion \& Conclusion}
\label{sec:conclusions}

This paper presented a large-scale analysis of the use of Web archiving services, such as archive.is and the Wayback Machine, on social media.
Our study is based on two data crawls: 1)~21M URLs obtained from the archive.is live feed, covering almost two years, and 
2)~356K archive.is plus 391K Wayback Machine URLs that were shared, over 14 months, on Reddit, Twitter, Gab, and 4chan's Politically Incorrect board (\dspol). 
We showed that these services are extensively used to archive and disseminate news, social network posts, and 
controversial content---in particular by users of fringe communities within Reddit and 4chan.
We also found that users not only rely on them to ensure persistence of Web content, but also to bypass certain censorship policies of some social networks.
Some subreddits, as well as 4chan's \dspol, actually nudge or force users to share archive URLs instead of direct link to news sources they perceive as having contrasting ideologies, taking away potentially hundreds of thousands of dollars in ad revenue.
Overall, our measurements highlight the importance of archiving services in the Web's information and ad ecosystems, and the need to carefully consider them when studying social media.

Our work also points to several open research avenues.
For instance, future work could better understand the role of archiving services in the dis/misinformation ecosystem, e.g., with respect to the content that gets archived and the context in which archive URLs are disseminated.
Moreover, further work could shed light on the actors archiving specific URLs in specific contexts, as well as how much traction they get on Web communities like Twitter and Reddit.
Finally, we believe that a deeper dive into the socio-technical and ethical implications of archiving services is warranted: they serve a crucial role in ensuring that Web content persists, but do so without regard to (and often in spite of) the rights and consent of content producers.

\descr{Acknowledgments.} This project has received funding from the European Union's Horizon 2020 Research and Innovation program under the Marie Sk\l{}odowska-Curie ENCASE project (GA No. 691025).
This work reflects only the authors' views and the Commission are not responsible for any use that may be made of the information it contains.

\small

\begin{thebibliography}{10}

\bibitem{ainsworth2011how}
S.~Ainsworth, A.~Alsum, H.~SalahEldeen, M.~C. Weigle, and M.~L. Nelson.
\newblock {How much of the web is archived?}
\newblock In {\em JCDL}, 2011.

\bibitem{alnoamany2014and}
Y.~Alnoamany, A.~Alsum, M.~C. Weigle, and M.~L. Nelson.
\newblock {Who and what links to the Internet Archive}.
\newblock {\em IJDL}, 2014.

\bibitem{alonso2017s}
O.~Alonso, V.~Kandylas, S.-E. Tremblay, J.~M. Hofman, and S.~Sen.
\newblock {What's Happening and What Happened: Searching the Social Web}.
\newblock In {\em WebSci}, 2017.

\bibitem{barford2014adscape}
P.~Barford, I.~Canadi, D.~Krushevskaja, Q.~Ma, and S.~Muthukrishnan.
\newblock {Adscape: harvesting and analyzing online display ads}.
\newblock In {\em WWW}, 2014.

\bibitem{gab_racism}
T.~Benson.
\newblock {Inside the “Twitter for racists”: Gab — the site where Milo
  Yiannopoulos goes to troll now}.
\newblock \url{https://goo.gl/Yqv4Ue}, 2016.

\bibitem{chatzakou2017hate}
D.~Chatzakou, N.~Kourtellis, J.~Blackburn, E.~{De Cristofaro}, G.~Stringhini,
  and A.~Vakali.
\newblock {Hate is not Binary: Studying Abusive Behavior of \#GamerGate on
  Twitter}.
\newblock In {\em HyperText}, 2017.

\bibitem{gdpr_right_forgotten}
{European Commission}.
\newblock {General Data Protection Regulation (GDPR), Art. 17}.
\newblock \url{https://gdpr-info.eu/art-17-gdpr/}, 2017.

\bibitem{georgiev2016gone}
M.~Georgiev and V.~Shmatikov.
\newblock {Gone in six characters: Short urls considered harmful for cloud
  services}.
\newblock {\em arXiv:1604.02734}, 2016.

\bibitem{gilbert2013underprovision}
E.~Gilbert.
\newblock {Widespread underprovision on Reddit}.
\newblock In {\em CSCW}, 2013.

\bibitem{glenski2017consumers}
M.~Glenski, C.~Pennycuff, and T.~Weninger.
\newblock {Consumers and Curators: Browsing and Voting Patterns on Reddit}.
\newblock {\em IEEE Transactions on Computational Social Systems}, 2017.

\bibitem{hackett2004accessibility}
S.~Hackett, B.~Parmanto, and X.~Zeng.
\newblock {Accessibility of Internet websites through time}.
\newblock In {\em ACM ASSETS}, 2004.

\bibitem{hale2017live}
S.~A. Hale, G.~Blank, and V.~D. Alexander.
\newblock {\em {Live versus Archive: Comparing a Web Archive and to a
  Population of Webpages}}.
\newblock UCL Press, 2017.

\bibitem{hine2017kek}
G.~E. Hine, J.~Onaolapo, E.~{De Cristofaro}, N.~Kourtellis, I.~Leontiadis,
  R.~Samaras, G.~Stringhini, and J.~Blackburn.
\newblock {Kek, Cucks, and God Emperor Trump: A Measurement Study of 4chan's
  Politically Incorrect Forum and Its Effects on the Web}.
\newblock In {\em ICWSM}, 2017.

\bibitem{holzmann2016dawn}
H.~Holzmann, W.~Nejdl, and A.~Anand.
\newblock {The Dawn of today's popular domains: A study of the archived German
  Web over 18 years}.
\newblock In {\em JCDL}, 2016.

\bibitem{guardian_reddit}
J.~Jackson.
\newblock {Moderators of pro-Trump Reddit group linked to fake news crackdown
  on posts}.
\newblock
  \url{https://www.theguardian.com/technology/2016/nov/22/moderators-trump-reddit-group-fake-news-crackdown},
  2016.

\bibitem{seth_rich}
{Jonah Engel Bromwich}.
\newblock {How the Murder of a D.N.C. Staff Member Fueled Conspiracy Theories}.
\newblock \url{https://nyti.ms/2rs9uGh}, 2017.

\bibitem{motherboard_archive_traffic}
J.~Koebler.
\newblock {Dear GamerGate: Please Stop Stealing Our Shit}.
\newblock
  \url{https://motherboard.vice.com/en_us/article/ypw5mj/dear-gamergate-please-stop-stealing-our-shit},
  2014.

\bibitem{koehler2004longitudinal}
W.~Koehler.
\newblock {A longitudinal study of Web pages continued: A consideration of
  document persistence}.
\newblock {\em Information Research}, 9(2), 2004.

\bibitem{lerner2017rewriting}
A.~Lerner, T.~Kohno, and F.~Roesner.
\newblock {Rewriting History: Changing the Archived Web from the Present}.
\newblock In {\em ACM CCS}, 2017.

\bibitem{lerner2016internet}
A.~Lerner, A.~K. Simpson, T.~Kohno, and F.~Roesner.
\newblock {Internet Jones and the Raiders of the Lost Trackers: An
  Archaeological Study of Web Tracking from 1996 to 2016}.
\newblock In {\em USENIX Security}, 2016.

\bibitem{weaponized_nytimes}
N.~MacFarquhar.
\newblock {A Powerful Russian Weapon: The Spread of False Stories}.
\newblock \url{https://nyti.ms/2k6880n}, 2016.

\bibitem{maggi2013two}
F.~Maggi, A.~Frossi, S.~Zanero, G.~Stringhini, B.~Stone-Gross, C.~Kruegel, and
  G.~Vigna.
\newblock {Two years of short URLs Internet measurement: Security threats and
  countermeasures}.
\newblock In {\em WWW}, 2013.

\bibitem{mondal2016forgetting}
M.~Mondal, J.~Messias, S.~Ghosh, K.~P. Gummadi, and A.~Kate.
\newblock {Forgetting in Social Media: Understanding and Controlling
  Longitudinal Exposure of Socially Shared Data}.
\newblock In {\em SOUPS}, 2016.

\bibitem{nikiforakis2012you}
N.~Nikiforakis, L.~Invernizzi, A.~Kapravelos, S.~Van~Acker, W.~Joosen,
  C.~Kruegel, F.~Piessens, and G.~Vigna.
\newblock {You are what you include: Large-scale evaluation of remote
  JavaScript inclusions}.
\newblock In {\em ACM CCS}, 2012.

\bibitem{nikiforakis2014stranger}
N.~Nikiforakis, F.~Maggi, G.~Stringhini, M.~Z. Rafique, W.~Joosen, C.~Kruegel,
  F.~Piessens, G.~Vigna, and S.~Zanero.
\newblock {Stranger danger: Exploring the ecosystem of ad-based URL shortening
  services}.
\newblock In {\em WWW}, 2014.

\bibitem{gab_hate_speech}
R.~Price.
\newblock {Google's app store has banned Gab, a social network popular with the
  far-right, for `hate speech'}.
\newblock
  \url{http://uk.businessinsider.com/google-app-store-gab-ban-hate-speech-2017-8},
  2017.

\bibitem{vice_banned_archive}
N.~Ralph.
\newblock {VICE Has Disabled Archiving Sites To Stop People Using Their Own
  Words Against Them}.
\newblock
  \url{http://theralphretort.com/vice-disabled-archiving-sites-against-them/},
  2017.

\bibitem{rivers2014ethical}
C.~M. Rivers and B.~L. Lewis.
\newblock {Ethical research standards in a world of big data}.
\newblock {\em F1000Research}, 2014.

\bibitem{soska2014automatically}
K.~Soska and N.~Christin.
\newblock {Automatically Detecting Vulnerable Websites Before They Turn
  Malicious}.
\newblock In {\em USENIX Security}, 2014.

\bibitem{bbc_4chan_pizzagate}
M.~Wendling.
\newblock {The saga of 'Pizzagate': The fake story that shows how conspiracy
  theories spread}.
\newblock \url{http://www.bbc.com/news/blogs-trending-38156985}, 2016.

\bibitem{gab_alt_right}
J.~Wilson.
\newblock {Gab: Alt-right's social media alternative attracts users banned from
  Twitter}.
\newblock
  \url{https://www.theguardian.com/media/2016/nov/17/gab-alt-right-social-media-twitter},
  2016.

\bibitem{zannettou2018gab}
S.~Zannettou, B.~Bradlyn, E.~De~Cristofaro, M.~Sirivianos, G.~Stringhini,
  H.~Kwak, and J.~Blackburn.
\newblock {What is Gab? A Bastion of Free Speech or an Alt-Right Echo Chamber?}
\newblock In {\em WWW Companion}, 2018.

\bibitem{zannettou2017web}
S.~Zannettou, T.~Caulfield, E.~{De Cristofaro}, N.~Kourtellis, I.~Leontiadis,
  M.~Sirivianos, G.~Stringhini, and J.~Blackburn.
\newblock {The Web Centipede: Understanding How Web Communities Influence Each
  Other Through the Lens of Mainstream and Alternative News Sources}.
\newblock In {\em ACM IMC}, 2017.

\end{thebibliography}

\bibliographystyle{abbrv}

\end{document}